\documentclass[a4paper,11pt]{article}
\usepackage{jheppub}
\usepackage[T1]{fontenc}
\usepackage{amssymb}
\usepackage{graphicx}
\usepackage{amsmath}
\usepackage{hyperref}
\usepackage{tensor}
\usepackage[utf8]{inputenc}
\usepackage{subfigure}
\usepackage{cases}
\usepackage{appendix}
\usepackage{multirow}

\def\be {\begin{equation}}
\def\ee {\end{equation}}
\def \nn {\nonumber}
\def \a {\alpha}
\def \b {\beta}

\def \r {\rho}

\def \d {\delta}

\def \e {\epsilon}

\def \m {\mu}
\def \n {\nu}
\def \dl{\Lambda_2}

\def \eff{\text{eff}}

\def \AdS{\text{AdS}}
\def \ads{\text{AdS}}

\def \pd{\partial}

\def \I{\mathcal{A}^{\text{WDW}}}

\def \bg{\bigg|}
\def \ff0{\Big(\frac{\f}{\f_0}\Big)}

\def \hna{\hat{n}_a}

\def \awdw{\mathcal{A}_{\WDW}}
\def \lctc{\mathcal{L}_\text{ct}^{\text{cha}}}
\def \lctn{\mathcal{L}_\text{ct}^{\text{neu}}}
\def \lct{\mathcal{L}_\text{ct}}
\def \rb{r_{\mathcal{B}}}
\def \rc{r_{\mathcal{U}}}
\def \rc{r_{\mathcal{U}}}
\def \M{\mathcal{M}}
\def \mo{\mathcal{O}}
\def \A{\mathcal{A}}
\def \B{\mathcal{B}}
\def \C{\mathcal{U}}

\def \S{\Sigma}

\def \f {\phi}
\def \fc{\f_{\mathcal{U}}}
\def \fb{\f_{\mathcal{B}}}

\def \fm{\f_\text{max}}
\def \fp{\f_+}
\def \fn{\f_-}

\def \tI{\widetilde{\A}}

\def \trf{t\rightarrow+\infty}
\def \rrf{r\rightarrow+\infty}

\def \fcm{\f_{c_{-}}}
\def \rcm{r_{c_{-}}}

\def \veff{V^{\text{eff}}(\f)}

\def \JT{\text{JT}}
\def \vth{V_\text{th}}
\def \ccb{c\bar{c}}
\def \E {\text{E}}
\def \surf{\text{surf}}
\def \WDW{\text{WDW}}
\def \GHY{\text{GHY}}
\def \bulk{\text{bulk}}
\def \joint{\text{joint}}

\def \bdy{\text{bdy}}
\def \ct{\text{ct}}

\def \oshl{\text{on-shell}}

\def \topo{\text{top}}
\def \total{\text{total}}
\def \neu{\text{neu}}
\def \cha{\text{cha}}

\title{\boldmath Revisit on holographic complexity in two-dimensional gravity}

\author[c]{Rong-Gen Cai,}
\author[a,b]{Song He,}
\author[d]{Shao-Jiang Wang,}
\author[a]{Yu-Xuan Zhang}

\affiliation[a]{Center for Theoretical Physics and College of Physics, Jilin University, Changchun 130012, People's Republic of China}
\affiliation[b]{Max Planck Institute for Gravitational Physics (Albert Einstein Institute), Am M\"uhlenberg 1, 14476 Golm, Germany}
\affiliation[c]{CAS Key Laboratory of Theoretical Physics, Institute of Theoretical Physics, Chinese Academy of Sciences, Beijing 100190, China}
\affiliation[d]{Tufts Institute of Cosmology, Department of Physics and Astronomy, Tufts University, 574 Boston Avenue, Medford, Massachusetts 02155, USA}

\emailAdd{cairg@itp.ac.cn}
\emailAdd{hesong@jlu.edu.cn}
\emailAdd{schwang@cosmos.phy.tufts.edu}
\emailAdd{yuxuanz18@mails.jlu.edu.cn}

\abstract{We revisit the late-time growth rate of various holographic complexity conjectures for neutral and charged AdS black holes with single or multiple horizons in two dimensional (2D) gravity like Jackiw-Teitelboim (JT) gravity and JT-like gravity. For complexity-action conjecture, we propose an alternative resolution to the vanishing growth rate at late-time for general 2D neutral black hole with multiple horizons as found in the previous studies for JT gravity. For complexity-volume conjectures, we obtain the generic forms of late-time growth rates in the context of extremal volume and Wheeler-DeWitt volume by appropriately accounting for the black hole thermodynamics in 2D gravity.}

\begin{document}
\maketitle
\flushbottom

\section{Introduction}\label{sec:introduction}

Despite the role as the earliest discovered fundamental interaction, gravity still remains as a myth by its quantum nature. General relativity reveals gravity as the manifestation of background geometry, and Ryu-Takayanagi formula \cite{Ryu:2006bv} motivates the pursuit of bulk geometry as a dual to some quantum information on the boundary, then a natural question to ask is whether gravity in the bulk could be reconstructed from quantum information on the boundary. This triangular relation between gravity, geometry and information is the main focus recently among the high-energy physics communities.

In the context of an eternal anti-de Sitter (AdS) black hole, quantum information on the boundary could be extracted from the dubbed thermo-field double (TFD) state \cite{Maldacena:2001kr}. However, entanglement entropy alone cannot capture all the quantum information on the boundary \cite{Hartman:2013qma,Susskind:2014moa},  nor can bulk geometry be fully reconstructed from the entanglement entropy alone \cite{Freivogel:2014lja}, hence there comes the need for complexity, which continues to growth even after reaching thermal equilibrium, similar to the growth of black hole interior. This holographic insight was first formulated in the context of complexity-volume (CV) conjecture \cite{Susskind:2014rva,Stanford:2014jda} (see also \cite{Couch:2016exn,Couch:2018phr} for other CV proposals) and later refined in the context of complexity-action (CA) conjecture \cite{Brown:2015bva,Brown:2015lvg} (see also \cite{Hayward:1993my,Brill:1994mb,Lehner:2016vdi,Ruan:2017tkr} for the clarification of action computation).

Although there have been  extensive studies on CV and CA conjectures from the bulk side (see, e.g. charged black holes \cite{Cai:2016xho,Cai:2017sjv}, UV divergences \cite{Chapman:2016hwi,Reynolds:2016rvl,Carmi:2016wjl,Kim:2017lrw}, subregion complexity \cite{Carmi:2016wjl,Ben-Ami:2016qex,Abt:2017pmf}, time-evolution \cite{Brown:2017jil,Carmi:2017jqz}, higher derivative gravities \cite{Cai:2016xho,Alishahiha:2017hwg,Cano:2018aqi}, Einstein-Maxwell-dilaton gravities \cite{Cai:2017sjv,Swingle:2017zcd,An:2018xhv,Alishahiha:2018tep}, Vaidya spacetimes \cite{Chapman:2018dem,Chapman:2018lsv}, switchback effect and quenches \cite{Susskind:2014jwa,Moosa:2017yvt}, and dS/FLRW boundaries \cite{Reynolds:2017lwq,An:2019opz}), the difficulty to establish a convincing holographic complexity lies in the lack of unique and well-defined notion for the complexity in the field theory from the boundary side \cite{Roberts:2016hpo,Hashimoto:2017fga,Kim:2017qrq,Chapman:2017rqy,Czech:2017ryf,Jefferson:2017sdb,Caputa:2017urj,Caputa:2017yrh,Bhattacharyya:2018bbv,Belin:2018bpg,Ali:2018fcz,Belin:2018fxe,Chapman:2018hou,Balasubramanian:2018hsu,Caputa:2018kdj,Guo:2018kzl,Yang:2018nda,Takayanagi:2018pml,Hackl:2018ptj,Khan:2018rzm,Yang:2018tpo,Bhattacharyya:2018wym,Camargo:2019isp,Bhattacharyya:2019kvj,Doroudiani:2019llj,Yang:2019udi,Sinamuli:2019utz,Caceres:2019pgf}. Note that the Lloyd bound \cite{Lloyd} on complexity growth rate was initially founded {in quantum mechanical system}, it makes the two-dimensional (2D) gravity \cite{Muta:1992xw,Nojiri:2000ja,Grumiller:2002nm}  be a special case to understand holographic complexity since the boundary theory could be a (super)conformal quantum mechanics (CQM)~\cite{Chamon:2011xk}.

Despite the fact that the AdS$_2$/CFT$_1$ is currently poor understood~\cite{Castro:2008ms,Cvetic:2016eiv}, the renewed interest on 2D gravity follows up the recent understanding~\cite{Polchinski:2016xgd,Maldacena:2016hyu} of Sachdev-Ye-Kitaev (SYK) model~\cite{Sachdev:1992fk,KitaevTalks}, which is conjectured to be dual to quantum gravity in two dimensions~\cite{Jensen:2016pah,Engelsoy:2016xyb}. As a toy model of the correspondences between bulk 2D gravity~\cite{Strominger:1998yg} and boundary quantum mechanics (QM)~\cite{Claus:1998ts}, the 2D AdS Jackiw-Teitelboim(JT) gravity~\cite{Jackiw:1982hg,Teitelboim:1983fg} is found in~\cite{Cadoni:2000gm} to be dual to a conformally invariant dynamics on the spacetime boundary that could be described in terms of a de Alfaro-Fubini-Furlan model~\cite{deAlfaro:1976vlx} coupled to an external source with conformal dimension two. Later in~\cite{Brigante:2002rv}, the asymptotic dynamics of 2D (A)dS JT gravity is further found to be dual to a generalized two-particle Calogero-Sutherland quantum mechanical model~\cite{Sutherland:1971kq,Gibbons:1998fa}. Based on JT model, the Almheiri-Polchinski (AP) model~\cite{Almheiri:2014cka} was recently introduced to study the back-reaction to AdS$_2$ since there are no finite energy excitations above the AdS$_2$ vacuum~\cite{Fiola:1994ir,Maldacena:1998uz}. A distinct feature of AP model is that the boundary time coordinate is lifted as a dynamical variable and could be described by the 1D Schwarzian derivative action~\cite{Maldacena:2016upp,Engelsoy:2016xyb}, of which the same pattern of action also appeares in the SYK model. This indicates that JT model might arise as a holographic description of infrared limit of SYK model.

The issue of holographic complexity for 2D JT gravity has been investigated in \cite{Brown:2018bms,Akhavan:2018wla,Alishahiha:2018swh,Goto:2018iay}. Naive computations of action growth rate for 2D JT gravity reduced from a near-extremal and near-horizon limit of Reissner-Nordstr$\ddot{\text{o}}$m (RN) black holes in higher dimensions was found to be perplexingly vanishing at late-time, of which, however, the late-time linear growth of the complexity could be restored by appending with an electromagnetic boundary term \cite{Brown:2018bms}  in order to ensure the correct sign of the dilaton potential during dimensional reduction. \cite{Akhavan:2018wla,Alishahiha:2018swh} proposed another restoration of the late-time linear growth of the complexity by appropriately relating the cut-off behind the horizon with the UV cut-off at the boundary (see also \cite{Hashemi:2019xeq}). Recently, \cite{Goto:2018iay} revealed an intriguing fact that the action growth rate for charged black hole is sensitive to the ratio between the electric and magnetic charges, and the previously identified vanishing result is due to the zero electric charged in the grand canonical ensemble, which could be dramatically changed by adding an appropriate surface term. In this paper, we would like to go beyond the 2D JT gravity.

{We investigated the late-time growth rate for neutral and charged asymptotically $\text{AdS}_2$ dilaton black hole solutions with single or multiple horizons in two dimensional (2D) gravity like Jackiw-Teitelboim (JT) gravity and JT-like gravity, by suing the  CA and CV conjectures. The main results are briefly summarized briefly\footnote{Table \ref{table}  presents the main results in a precise way.}:
  i)  In the case with  CA conjecture, for charged black holes with single or multiple horizons and neutral black holes with single horizon, the late-time growth rate of complexity exactly saturates the charged or neutral versions of Lloyd bound, respectively. In particular, for neutral black hole with multiple horizons, the late-time growth rate is found to be vanished if the conventional approach is naively adopted. To resolve this puzzle, we make use of magnetic/electrical duality of 4D RN AdS black hole to restore the late-time linear growth of complexity, which can be regarded as a generalization of the method introduced in \cite{Brown:2018bms}.
ii) In the CV1.0 conjecture, the universal form of late-time growth rate is obtained for generic neutral and charged asymptotically $\text{AdS}_2$ dilaton black hole solutions with single or multiple horizons. In the CV 2.0 version, we obtain the generic forms of late-time growth rates in the context of Wheeler-de Witt volume by appropriately accounting for the 2D black hole thermodynamics.}

The organization of the remaining parts of this paper is as follows. The holographic complexity in terms of CA, CV 1.0 as well as CV 2.0, in 2D neutral and charged black holes have been investigated in Sec. \ref{sec:CA},  Sec. \ref{sec:CV1.0} and Sec. \ref{sec:CV2.0}, respectively. Sec. \ref{sec:conclusion} is devoted to conclusions. The appendix \ref{app:A} elaborates the form of counter term and topological term for the 2D black holes.

\section{CA}\label{sec:CA}

In this section, we evaluate the late-time growth rate of holographic complexity for neutral and charged eternal $\AdS_2$ black holes using the CA conjecture, which claims that the complexity of a TFD state living on the boundaries is proportional to the gravitational action evaluated on the Wheeler-DeWitt (WDW) patch
\begin{align}\label{eq:CA}
\mathcal{C_A}=\frac{\mathcal{A}_\mathrm{WDW}}{\pi\hbar}.
\end{align}
Here the coefficient $1/\pi\hbar$ is chosen in such a way that the late-time limit $t\equiv t_L+t_R\to\infty$ of $\mathcal{C_A}$ growth rate exactly saturates the Lloyd bound \cite{Lloyd}
\begin{align}\label{eq:Lloyd}
\left.\frac{\mathrm{d}\mathcal{C_{A}}}{\mathrm{d}t}\right|_{t\to\infty}=\frac{2M}{\pi\hbar}
\end{align}
at least for AdS Schwarzschild black hole in $D\geq4$ dimensions in Einstein gravity \cite{Brown:2015lvg}, beyond which various corrections to $2M$ at late-time are expected for other neutral AdS black holes also with a single horizon. However, for charged AdS black holes with both inner and outer horizons, the holographic complexity given by CA conjecture saturates at late-time a different form \cite{Cai:2016xho} as
\begin{align}\label{eq:Lloyd2}
\left.\frac{\mathrm{d}\mathcal{C_A}}{\mathrm{d}t}\right|_{t\to\infty}=\frac{1}{\pi\hbar}(M-\mu Q-\Omega J)\bigg|_{r_-}^{r_+},
\end{align}
where the inner horizon $r_-$  emerges  besides the outermost horizon $r_+$ due to the presence of conserved charges $Q$ and $J$ with corresponding chemical
potentials  $\mu$ and $\Omega$, respectively. The form \eqref{eq:Lloyd2} (see also \cite{Huang:2016fks,Liu:2019mxz}) is quite generic for charged AdS black holes with double horizons even beyond the Einstein gravity. Nevertheless, there also exists other special cases of neutral black holes with multiple horizons and charged black holes with a single horizon (for example, see \cite{Cai:2017sjv}). All these cases mentioned above will be studied below for 2D AdS black holes beyond simple JT gravity.

\subsection{Neutral black holes}\label{sec:CAneutral}

We start  with the  neutral AdS black holes in 2D gravity with Einstein-Hilbert-dilaton action of form
\begin{align}\label{eq:CAneutral}
\mathcal{A}[g,\phi]
=\frac{1}{2G}\int_\mathcal{M}\mathrm{d}^2x\sqrt{-g}\left(\phi R(g)+\frac{V(\phi)}{L^2}\right)
+\frac{1}{G}\int_{\partial\mathcal{M}}\mathrm{d}x\sqrt{-h}\Big(\phi K-\mathcal{L}_\mathrm{ct}^\mathrm{neu}(\phi)\Big),
\end{align}
where $\mathcal{L}_\mathrm{ct}^\mathrm{neu}$ is a boundary counter term that renders the Euclidean on-shell action finite. With the ansatz of a linear dilaton of mass scale $\alpha$ and Schwarzschild gauge of the metric
\begin{align}
\phi&=\alpha r, \label{eq:phir}\\
\mathrm{d}s^2&=-f(\phi)\mathrm{d}t^2+f(\phi)^{-1}\mathrm{d}r^2,
\end{align}
the corresponding equations-of-motion (EOMs)
\begin{align}\label{eq:CAneutralEOM}
R&=-\frac{V'(\phi)}{L^2},\\
0&=-\nabla_\mu\nabla_\nu\phi+g_{\mu\nu}\left(\nabla^2\phi-\frac{V(\phi)}{2L^2}\right)
\end{align}
could be solved as
\begin{align}
f(\phi)&=-\frac{2GM}{\alpha}+j(\phi), \label{eq:fphi}\\
j(\phi)&=\frac{1}{\a^2L^2}\int^\f V(\f')d\f',\label{eqj}
\end{align}
where $M$ is the ADM mass as shown in Appendix \ref{app:A} in order to preserve the thermodynamic relation with the black hole temperature and Wald entropy  \cite{Wald:1993nt} defined respectively by
\begin{align}
T&=\frac{f'(r_+)}{4\pi}=\frac{V(\phi_+)}{4\pi\alpha L^2},\label{eq:CAneutralT}\\
S&=\frac{2\pi\phi_+}{G},\quad \phi_+\equiv\alpha r_+.
\end{align}
The boundary term is therefore computed by

\begin{align}
\sqrt{-h}K&=\frac{V(\phi)}{2\alpha L^2}.\\
\mathcal{L}_\mathrm{ct}^\mathrm{neu}&=\alpha\sqrt{j(\phi)},\label{A.4}
\end{align} One could refer to (\ref{A.4}) in Appendix \ref{app:A} for more details.

We next turn to  a particular form of potential motivated by \cite{Kettner:2004aw}
\begin{align}\label{eq:CAneutralV}
V(\phi)=2\phi+B+\sum\limits_iB_i\phi^{b_i},\quad b_i\neq0,1
\end{align}
which becomes JT gravity when $B=B_i=0$. The corresponding on-shell Ricci scalar curvature $R$ is
\begin{align}
R=-\a^2f''(\f)=-\frac{2}{L^2}+\sum_{i}b_iB_i\f^{b_i-1},
\end{align}
where a curvature singularity would appear at $r=0$ when there is a nonzero $B_i$. What's more, one should restrict $b_i<1$ to obtain an asymptotically $\mathrm{AdS}_2$ geometry. Based on above arguments, the black hole solutions generated by \eqref{eq:CAneutralV} could be divided into two classes: the one with single horizon and the one with multiple horizons, which will be studied in detail below.

\subsubsection{Single horizon}\label{subsubsec:CAneutralsingle}

\begin{figure}
\centering
\includegraphics[width=0.46\textwidth]{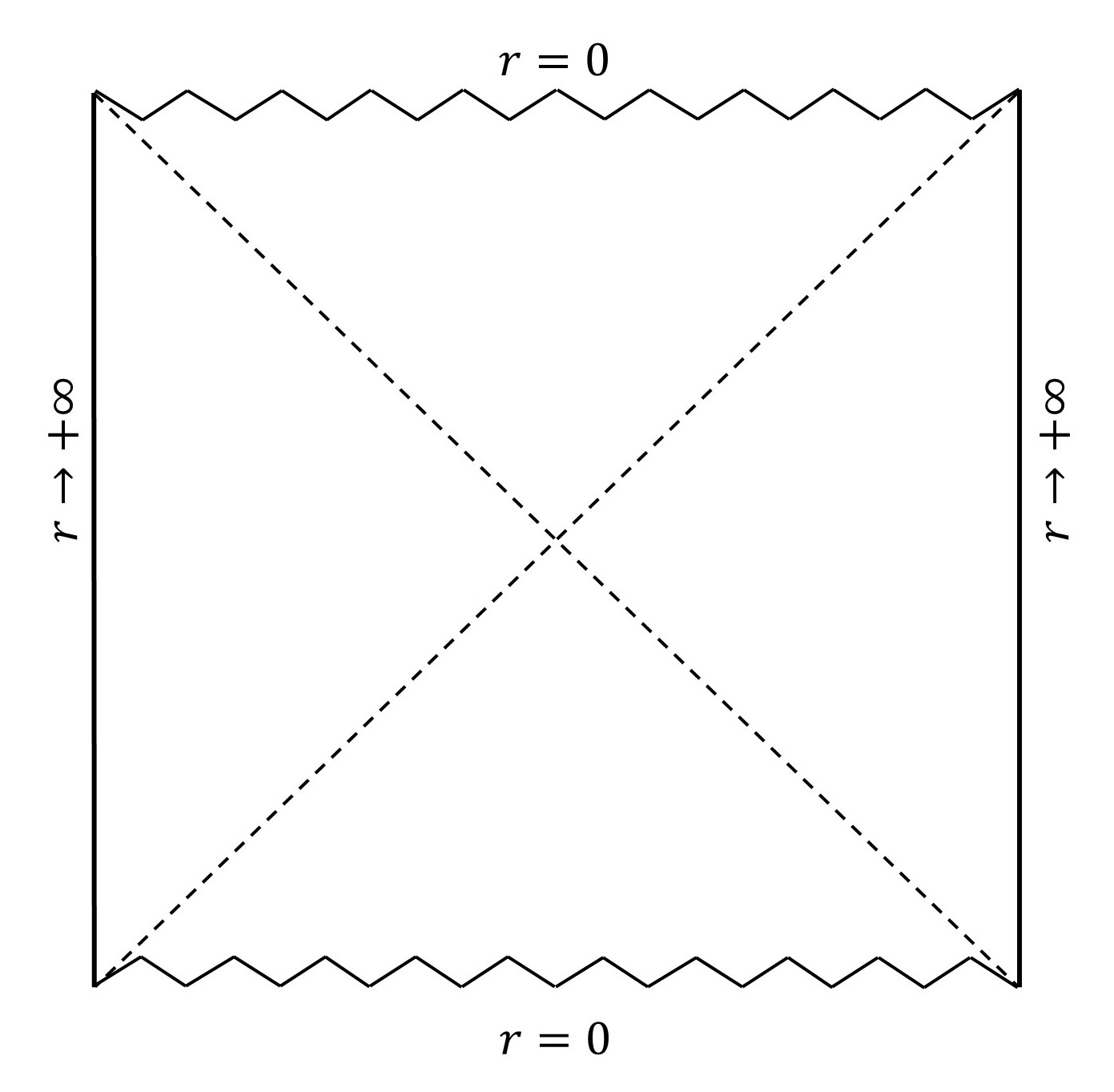}
\includegraphics[width=0.52\textwidth]{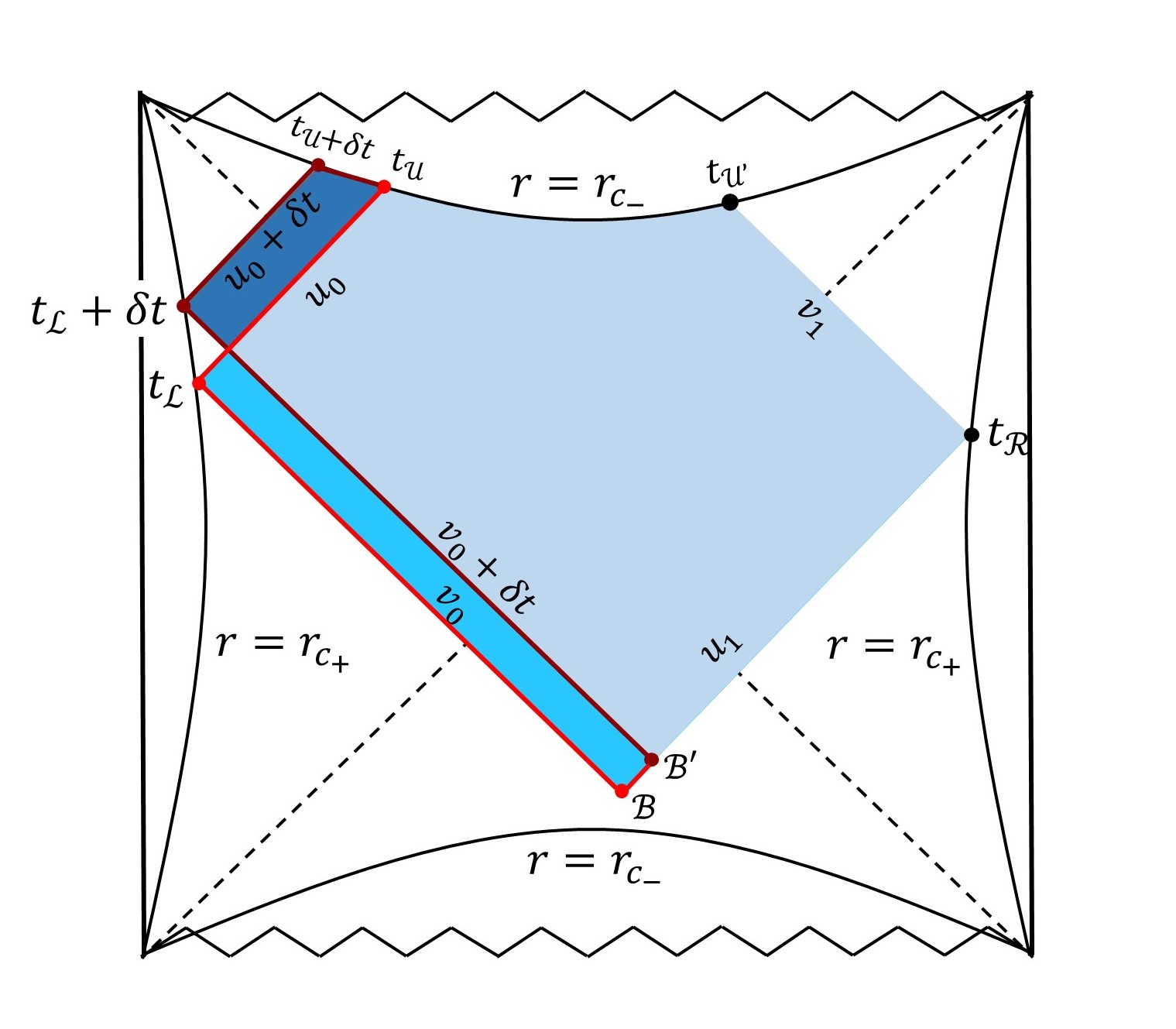}\\
\caption{\textit{Left panel}\label{fig:CAneutralsingle}: Neutral (Charged) AdS black hole in 2D gravity with a single horizon at $r=r_h$ (dashed lines),  spatial singularities at $r=0$ (wavy lines) and asymptotic boundaries at $r\to\infty$ (vertical lines). \textit{Right panel}: The change of WDW patches from the one anchored on $t_L$ and $t_R$ (pale blue and bright blue regions) to the one anchored on $t_L+\delta t$ and $t_R$ (pale blue and dark blue regions). The UV and IR cutoffs are placed at $r=r_{c_+}$ and $r=r_{c_-}$, respectively.}
\end{figure}

If there is a nonzero $B_i$, then the non-extreme single horizon solutions are allowed, namely, $f(\phi)=0$ has one positive root\footnote{e.g.,$V(\phi)=2\phi+\frac{1}{\phi}$, then $f(\phi)=\frac{-2GM}{\a}+\frac{1}{\a^2L^2}\big(\phi^2+\log\phi\big)$.}, of which the Penrose diagrams share the same feature as shown in Fig.\ref{fig:CAneutralsingle}. Without lost of generality, the left and right boundaries could be related to the Schwarzschild time by $t_L=t, t_R=-t$, and the Eddington-Finkelstein coordinates $u=t+r^*$ and $v=t-r^*$ with $r^*=\int f(r)^{-1}\mathrm{d}r$ and $\rho(r^*(r))=r$ are introduced to rewrite the metric as $\mathrm{d}s^2=-f(r)\mathrm{d}u^2+2\mathrm{d}u\mathrm{d}r=-f(r)\mathrm{d}v^2-2\mathrm{d}v\mathrm{d}r$.

Following the convention of \cite{Lehner:2016vdi}, the total change of action due to  changing the WDW patch reads
\begin{align}
\d \A_{\WDW}=\d\I_{\bulk}+\d\I_{\text{ren.}\surf}+\d\I_{\joint},
\end{align}
where $\A^{\WDW}_{\text{ren.}\surf}$ is the surface term we renormalized to make the late-time action growth rate converge,
\begin{align}
A^{\WDW}_{\text{ren.surf}}=\int_{r=r_{c_-}\bigcap\WDW{\text{patch}}}\sqrt{|h|}\mathrm{d}t\Big(\phi K-\frac{\a j(\f)}{\sqrt{|f(\f)|}}\Big).
\end{align}
The bulk term contribution is
\begin{align}
\d\A^{\WDW}_{\bulk}&=\frac{1}{2G}\int^{u_0+\d t}_{u_0}\mathrm{d}u\int_{\rcm}^{\rho\big(\frac{u-(v_0+\d t)}{2}\big)}\frac{V(\phi)-\phi V'(\phi)}{L^2}\mathrm{d}r\nn\\
&-\frac{1}{2G}\int_{v_0}^{v_0+\d t}\mathrm{d}v\int_{\rho\big(\frac{u_1-v}{2}\big)}^{\rho\big(\frac{u_0-v}{2}\big)}\frac{V(\phi)-\phi V'(\phi)}{L^2}\mathrm{d}r\nn\\
&=\d t\Big(\frac{\a j(\f)}{G}-\frac{\f V(\f)}{2G\a L^2}\Big)\bg^{\fb}_{\fcm},\label{bulk}
\end{align}
the contribution of the renormalised-surface term is

\begin{align}
\d\A^{\WDW}_{\text{ren.}\surf}&=-\frac{1}{G}\int_{u_0-r^*(r_{c_-})}^{u_0-r^{*}(r_{c_-})+\d t}\sqrt{|h|}\mathrm{d}t\Big(\phi K-\frac{\a j(\f)}{\sqrt{|f(\f)|}}\Big)\nn\\
&=\d t\Big(-\frac{\f V(\f)}{2G\a L^2}+\frac{\a j(\f)}{G}\Big)\bg_{\f=\f_{c_{-}}},\label{GHY}
\end{align}
and the joint term contribution is
\begin{align}
\d\A^{\WDW}_{\joint}&=\frac{1}{G}\phi_{\mathcal{B}'}a_{\mathcal{B}'}-\frac{1}{G}\phi_{\mathcal{B}}a_{\mathcal{B}}\nn\\
&=\d t\bigg(\frac{\phi V(\f)}{2G\a L^2} +\frac{\a f(\f)}{2G}\log\Big|\frac{f(\phi)}{c\bar{c}}\Big|\bigg)\bg_{\f=\fb}\label{joint},\end{align}
then the total variation of $\A_{\WDW}$ is

\begin{align}
\d \A_{\WDW}=\d t\Big(\frac{\a j(\f)}{G}+\frac{\a f(\f)}{2G}\log\Big|\frac{f(\f)}{\ccb}\Big|\Big)\bg_{\f=\fb}\label{dyn-total}.
\end{align}

As the boundary time goes to infinity, $\phi_{\mathcal{B}}\rightarrow\phi_h$, then we have

\begin{align}
\frac{\mathrm{d}\A_{\WDW}}{\mathrm{d}t}\bg_{t\rightarrow+\infty}=2M\label{s1e3},
\end{align}
which is the same as that of Schwartzchild-AdS black hole in higher dimensional gravity \cite{Brown:2015lvg}. Note that the extra boundary counter term we introduced in \eqref{GHY} not only offsets the possible divergence\footnote{Without the counter term, the late-time action growth rate will diverge when $ b_i\leq-1$ in \eqref{eq:CAneutralV}.}, but also makes the action growth rate saturate the desired Lloyd bound. It's also important to point out that the form of the additional  counter term is not uniquely fixed.

Finally, the rate (2.23) only depends on the mass. To see how the rate depends on the entropy, one can rewrite $ M$ as a function of entropy $S$ and temperature $T$ which are determined by the black hole horizon $r_+$, that is
\begin{align}
M=ST\Big(\frac{\int^{\phi_+}V(\phi)\mathrm{d}\phi}{\phi_+ V(\phi_+)}\Big)\Big|_{\phi_+=\frac{GS}{2\pi}}.
\end{align}
For generic form of the dilaton potential (2.15), the result is
\begin{align}
M=ST\cdot\frac{1+B\phi_+^{-1}+\sum_i\frac{B_i}{b_i+1}\phi_+^{b_i-1}}{2+B\phi_+^{-1}+\sum_jB_j\phi_+^{b_j-1}}\Bigg|_{\phi_+=\frac{GS}{2\pi}},~~~(b_i\neq-1).\label{3}
\end{align}
Note that when $B=B_i=0$, i.e., JT gravity, $M=\frac{1}{2}ST$, which is consistent with the result shown in [91]. One may reduce \eqref{3} to an unified form  in the case of massive black holes
\begin{align}
M\approx\frac{1}{2}ST.
\end{align}

\subsubsection{Multiple horizons}\label{universal}

\begin{figure}
\centering
\includegraphics[width=0.49\textwidth]{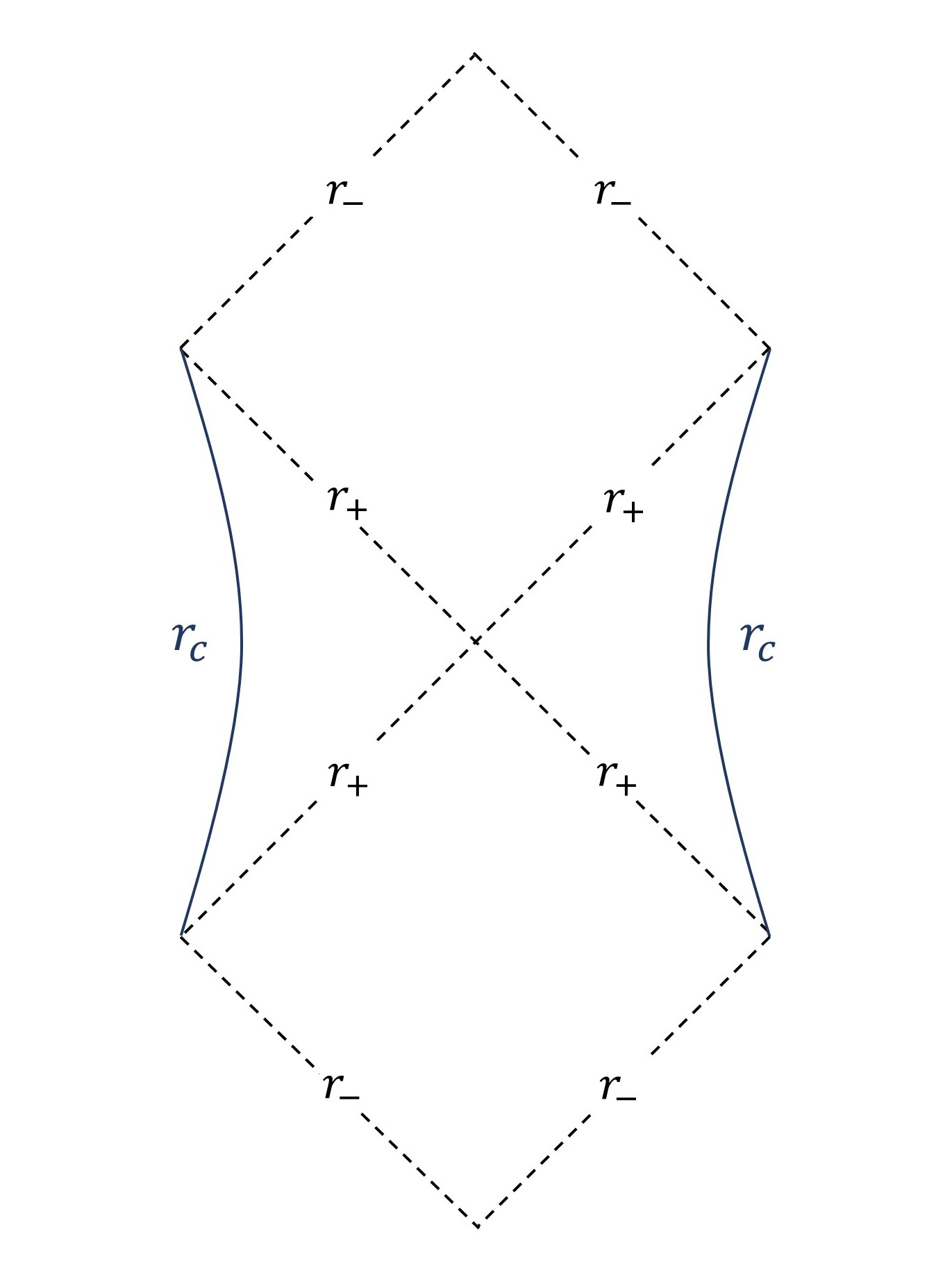}
\includegraphics[width=0.46\textwidth]{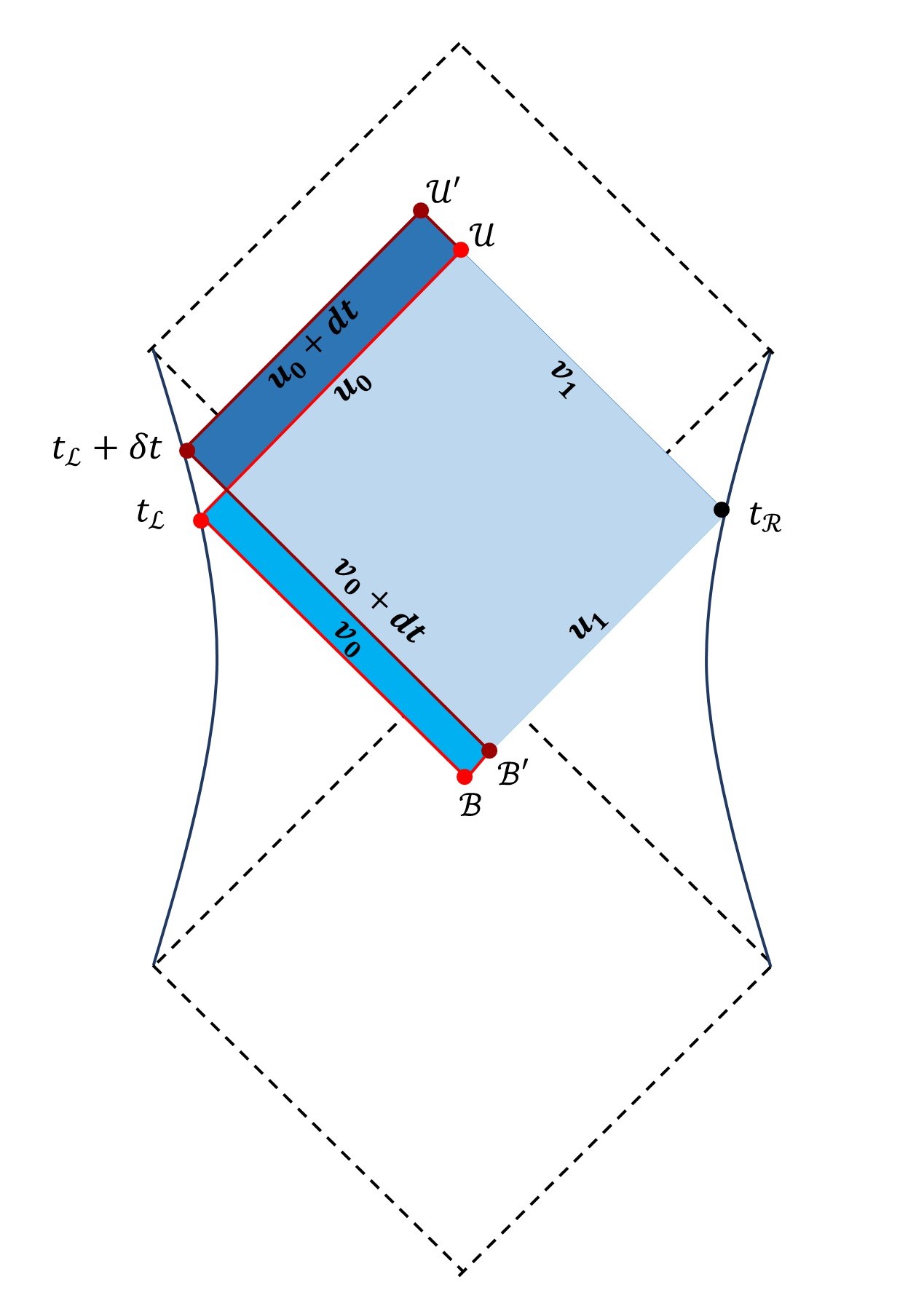}\\
\caption{\textit{Left panel}\label{fig:CAneutraldouble}: The universal feature shared by the Penrose diagrams of $\AdS_2$ black holes with multiple horizons (dashed lines), where $r_\pm$ are the outermost and next-to-outermost horizons, respectively, and $r_c$ corresponds to the UV-cutoff. \textit{Right panel}: The change of WDW patches from the one anchored on $t_L$ and $t_R$ (pale blue and bright blue regions) to the one anchored on $t_L+\delta t$ and $t_R$ (pale blue and dark blue regions).}
\end{figure}

For neutral AdS black hole in 2D gravity with multiple horizons, the Penrose diagram shares the same feature as shown in Fig. \ref{fig:CAneutraldouble}. Now the total change of action due to the change of WDW patch reads
\begin{align}
\d\A_{\WDW}=\d \A^{\WDW}_{\bulk}+\d\A^{\WDW}_{\surf}+\d\A^{\WDW}_{\joint},
\end{align}
where the null surface terms can be made vanish with the choice of affine parameter for the generator of null surfaces, while the bulk and joint contributions are evaluated as
\begin{align}
\d\A^{\WDW}_{\bulk}&=\frac{1}{2G}\int^{u_0+\d t}_{u_0}\mathrm{d}u\int^{\rho(\frac{u-v_0-\d t}{2})}_{\rho(\frac{u-v_1}{2})}\big(-\frac{\phi V'(\phi)}{L^2}+\frac{V(\phi)}{L^2}\big)\mathrm{d}r\nn\\
&-\frac{1}{2G}\int_{v_0}^{v_0+\d t}\mathrm{d}v\int_{\rho(\frac{u_1-v}{2})}^{\rho(\frac{u_0-v}{2})}\big(-\frac{\phi V'(\phi)}{L^2}+\frac{V(\phi)}{L^2}\big)\mathrm{d}r\nn\\
&=\d t\frac{\a j(\phi)}{G}\bigg|_{\fc}^{\fb}-\d t\frac{\phi V(\phi)}{2G\a L^2}\bigg|_{\fc}^{\fb}~~,
\end{align}
and
\begin{align}
\d\A^{\WDW}_{\joint}&=\frac{1}{G}\big(\phi_{\mathcal{B}'}a_{\mathcal{B}'}-\phi_{\mathcal{B}}a_{\mathcal{B}}\big)+
\frac{1}{G}\big(\phi_{\mathcal{U}'}a_{\mathcal{U}'}-\phi_{\mathcal{U}}a_{\mathcal{U}}\big)\nn\\
&=\frac{\d t}{G}\Bigg[\frac{\a f(r)\log|\frac{f(r)}{\ccb}|}{2}\bigg|_{\rc}^{\rb}+\frac{\phi V(\phi)}{2\a^2L^2}\bigg|^{\fb}_{\fc}\Bigg]~,
\end{align}
respectively. Now the growth rate of total action at late-time for neutral $\mathrm{AdS}_2$ black holes with multiple horizons vanishes similarly as JT gravity \cite{Brown:2018bms,Akhavan:2018wla,Alishahiha:2018swh,Goto:2018iay}
\begin{align}\label{CA0}
\frac{\mathrm{d}\A_{\WDW}}{\mathrm{d}t}\bg_{\trf}=\Big(\frac{\mathrm{d}\A^{\WDW}_{\bulk}}{\mathrm{d}t}+\frac{\mathrm{d}\A_{\joint}^{\WDW}}{\mathrm{d}t}\Big)\bg_{\trf}=\frac{\a j(\f)}{G}\bg_{\fn}^{\fp}=0,
\end{align}
which will be remedied below in Sec.\ref{subsec:resolution} with similar methods as proposed in \cite{Brown:2018bms,Akhavan:2018wla,Alishahiha:2018swh} as well as our new treatment from a dual charged black hole.

\subsection{Charged black holes}\label{charg}

We next move to the charged AdS black holes in 2D gravity with Einstein-Maxwell-dilaton action of form
\begin{align}
\A&=\frac{1}{2G}\int_{\mathcal{M}}\bigg(\phi R+\frac{V(\phi)}{L^{2}}-\frac{G}{2}W(\phi)F^{2}\bigg)\sqrt{-g}\mathrm{d}^2x+\frac{1}{G}\int_{\pd\mathcal{M}}\bigg(\phi K-\lctc(\f)\bigg)\sqrt{-h}\mathrm{d}x~,
\end{align}
where $\lctc$ is the counter term for charged black holes that renders the Euclidean on-shell action finite.
With the ansatz of a linear dilaton and RN gauge
\begin{align}
\phi(r,t)=\a r,~~\mathrm{d}s^2=-f(\f)\mathrm{d}t^2+f(\f)^{-1}\mathrm{d}r^2,
\end{align}
the corresponding EOMs
\begin{align}
R+\frac{1}{L^2}V'(\phi)-\frac{G}{2}W'(\phi)F^2&=0\label{R-charged}~,\\
\nabla_{\m}\nabla_{\n}\phi-g_{\m\n}\bigg(\nabla^{2}\phi-\frac{V(\phi)}{2L^2}\bigg)-\frac{GW(\phi)}{4}g_{\m\n}F^2+GW(\phi)F_{\m\r}F_{\n}^{~\r} &=0~,\\
\nabla_{\m}\big(W(\phi)F^{\m\n}\big)&=0~,\label{a1e1}
\end{align}
could be solved as
\begin{align}
A_t&=\frac{Qk(\f)}{\a},~~F_{tr}=-\frac{Q}{W(\phi)},~~F_{\m\n}F^{\m\n}=-\frac{2Q^2}{W(\phi)^2}\label{FmnFmn},\\
f(\f)&=-\frac{2GM}{\a}+j(\phi)-\frac{GQ^2}{\a^2}k(\phi)\label{re13},\\
j(\phi)&=\int^{\f} \frac{V(\phi')}{\a^2L^2}\mathrm{d}\phi',~~k(\phi)=\int^{\f}\frac{1}{W(\phi')}\mathrm{d}\phi'\label{er3},
\end{align}
where $Q$ is the electric charge of the system. Similar to the neutral case, the potential is taken as $V(\phi)=2\phi+B+\sum\limits_iB_i\phi^{b_i}$ with $b_i\neq0, 1$. If one further specifies $W(\phi)=A\phi^a$,  then the Ricci scalar
\begin{align}
R=-\frac{2}{L^2}-\frac{1}{L^2}\sum_{i}b_iB_i\f^{b_i-1}-\frac{aGQ^2}{A}\f^{-1-a}\label{Rcharge}
\end{align}
exhibits an asymptotically $\AdS_2$ boundary provided that $0\neq b_i<1$ and $a\geq-1$. A curvature singularity arises at $\f=0$ when $\exists$ $B_i\neq0$ or $a>-1$.

\subsubsection{Single horizon}

The Penrose diagram for charged $\mathrm{AdS}_2$ black hole with a single horizon (non-extreme) is the same as the neutral case in Fig. \ref{fig:CAneutralsingle}, of which the change in WDW patch leads to similar change in action as
\begin{align}
\d \A_{\WDW}=\d\A^{\WDW}_{\bulk}+\d\A_{\text{ren.}\surf}^{\WDW}+\d\A^{\WDW}_{\joint},
\nn\end{align}
like the case of neutral black hole, the surface term would be renormalized to obtain a reasonable result. The contribution from the bulk term is
\begin{align}
\d\I_{\bulk}&=\frac{1}{2G}\int^{u_0+\d t}_{u_0}\mathrm{d}u\int^{\rho(\frac{u-v_0-\d t}{2})}_{\rho(\frac{u-v_1}{2})}\big(\f R+\frac{V(\f)}{L^2}-\frac{G}{2}W(\f)F^2\big)\mathrm{d}r\nn\\
&-\frac{1}{2G}\int_{v_0}^{v_0+\d t}\mathrm{d}v\int_{\rho(\frac{u_1-v}{2})}^{\rho(\frac{u_0-v}{2})}\big(\f R+\frac{V(\f)}{L^2}-\frac{G}{2}W(\f)F^2\big)\mathrm{d}r\nn\\
&=\d t\bigg(\frac{\a j(\f)}{G}-\frac{\f V(\f)}{2G\a L^2}+\frac{Q^2\f}{2\a W(\f)}\bigg)\bg^{\fb}_{\fcm},\end{align}
the contribution from the renormalized-surface term is

\begin{align}
\d\I_{\text{ren.}\surf}&=-\frac{1}{G}\int_{u_0-r^*(r_{c_-})}^{u_0-r^*(r_{c_-})+\d t}\sqrt{|h|}\mathrm{d}t\Big(\phi K-\frac{\a j(\f)}{\sqrt{|f(\f)|}}\Big)\nn\\
&=\d t\bigg(\frac{-\f V(\f)}{2G\a L^2}+\frac{\f Q^2}{2\a W(\f)}+\frac{\a j(\f)}{G}\bigg)\bg_{\f=\fcm},
\end{align}

and the contribution from the joint term is
\begin{align}
\d\I_{\joint}&=\frac{1}{G}(\f_{\B'}a_{\B'}-\fb a_{\B})\nn\\
&=\d t\bigg(\frac{\f V(\f)}{2G\a L^2}-\frac{Q^2\f}{2\a W(\f)}+\frac{\a f(\phi_{\mathcal{B}})}{2G}\log\Big|\frac{f(\phi_{\mathcal{B}})}{c\bar{c}}\Big|\bigg)\bg_{\f=\f_\B}.
\end{align}
Hence the total variation of $\awdw$ is

\begin{align}
\d\A_{\WDW}&=\d t\Big(\frac{\a j(\f)}{G}+\frac{\a f(\phi)}{2G}\log\Big|\frac{f(\phi)}{c\bar{c}}\Big|\Big)\bg_{\f=\fb},
\end{align}

whose growth rate at late-time reads

\begin{align}
\frac{\mathrm{d}\A_{\WDW}}{\mathrm{d}t}\bg_{\trf}&=2M-\m Q,\label{charge1h}
\end{align}
where $\m$ is the chemical potential of the charged black hole, see Appendix \ref{app.cha} for more details. One can see that when $Q\rightarrow0$, the action growth goes back to the neutral case as expected.
\subsubsection{Multiple horizons}\label{double-chagre}

The Penrose diagram for charged $\mathrm{AdS}_2$ black hole with multiple horizons shares the same features as the neutral case in Fig. \ref{fig:CAneutraldouble}, of which the change in WDW patch leads to similar change in action as
\begin{align}
\d \A_{\WDW}=\A^{\WDW}_{\text{dark blue\&red}}-\A^{\WDW}_{\text{bright blue\&red}}=\d \A_{\bulk}+\d \A_{\joint},\nn
\end{align}
where the bulk contribution from Einstein-dilaton part is
\begin{align}
\d\I_{\bulk_1}&=\frac{1}{2G}\int_{\d v}\bigg(\phi R+\frac{V}{L^2}\bigg)\sqrt{-g}\mathrm{d}^2x\nn\\
&=\d t\bigg(\frac{\a j(\phi)}{G}-\frac{\phi V(\phi)}{2G\a L^2}-\frac{Q^2k(\phi)}{2\a}+\frac{Q^2\phi}{2\a W(\phi)}\bigg)\Bigg|^{\fb}_{\fc},
\end{align}
and the bulk contribution from Einstein-Maxwell part is
\begin{align}
\d \I_{\bulk_2}&=-\frac{1}{4}\int_{\d v}W(\phi)F^2\sqrt{-g}\mathrm{d}^2x\nn\\
&=\d t\frac{Q^2k(\phi)}{2\a}\bigg|^{\fb}_{\fc}.
\end{align}
The contribution from the joint term is
\begin{align}
\d\I_{\joint}&=\frac{1}{G}\big(\phi_{\B'}a_{\B'}-\phi_\B a_\B\big)+\frac{1}{G}\big(\phi_{\C'}a_{\C'}-\phi_\C a_\C\big)\nn\\
&=\d t\bigg(\frac{\phi V(\phi)}{2G\a L^2}-\frac{Q^2\phi}{2\a W(\phi)}\bigg)\bigg|^{\fb}_{\fc}+\d t\frac{f(r)\log\frac{-f(r)}{\ccb}}{2G}\bigg|_{\rc}^{\rb}.
\end{align}
Hence the total variation is
\begin{align}
\d\A_{\WDW}&=\d\I_{\bulk_1}+\d\I_{\bulk_2}+\d\I_{\joint}\nn\\
&=\d t\frac{\a j(\phi)}{G}\bigg|^{\fb}_{\fc}+\d t\frac{f(r)\log\frac{-f(r)}{\ccb}}{2G}\bigg|_{\rc}^{\rb},
\end{align}
whose growth rate at late-time reads
\begin{align}\label{CAmultiplehorizon}
\frac{\mathrm{d}\A_{\WDW}}{\mathrm{d}t}\bg_{\trf}=\frac{\a j(\phi)}{G}\bigg|_{\f_-}^{\f_+}=\m_{-}Q-\m_{+}Q~.
\end{align}

\subsection{Resolutions for vanishing late-time growth rate}\label{subsec:resolution}

\begin{figure}
\centering
\includegraphics[width=0.49\textwidth]{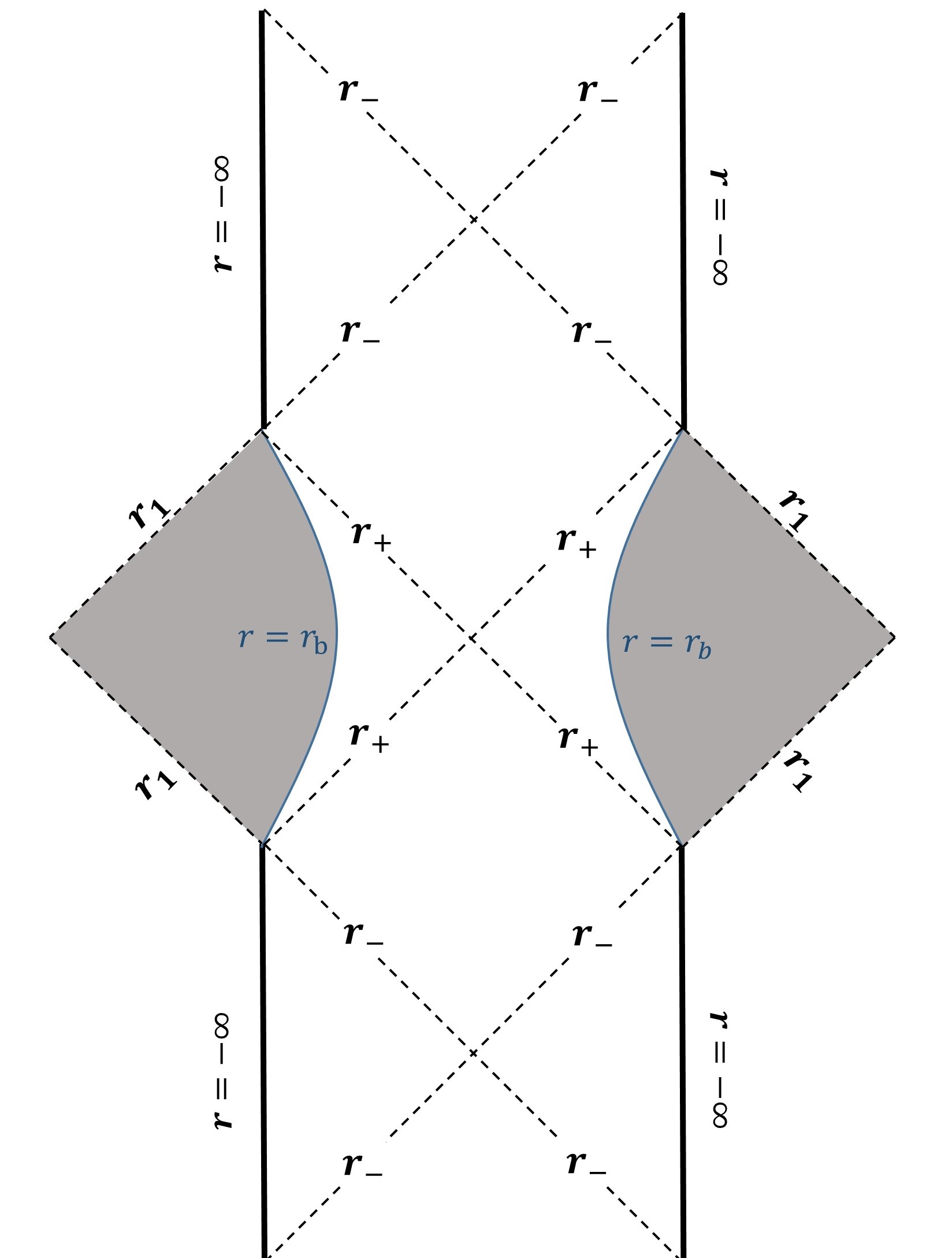}
\includegraphics[width=0.49\textwidth]{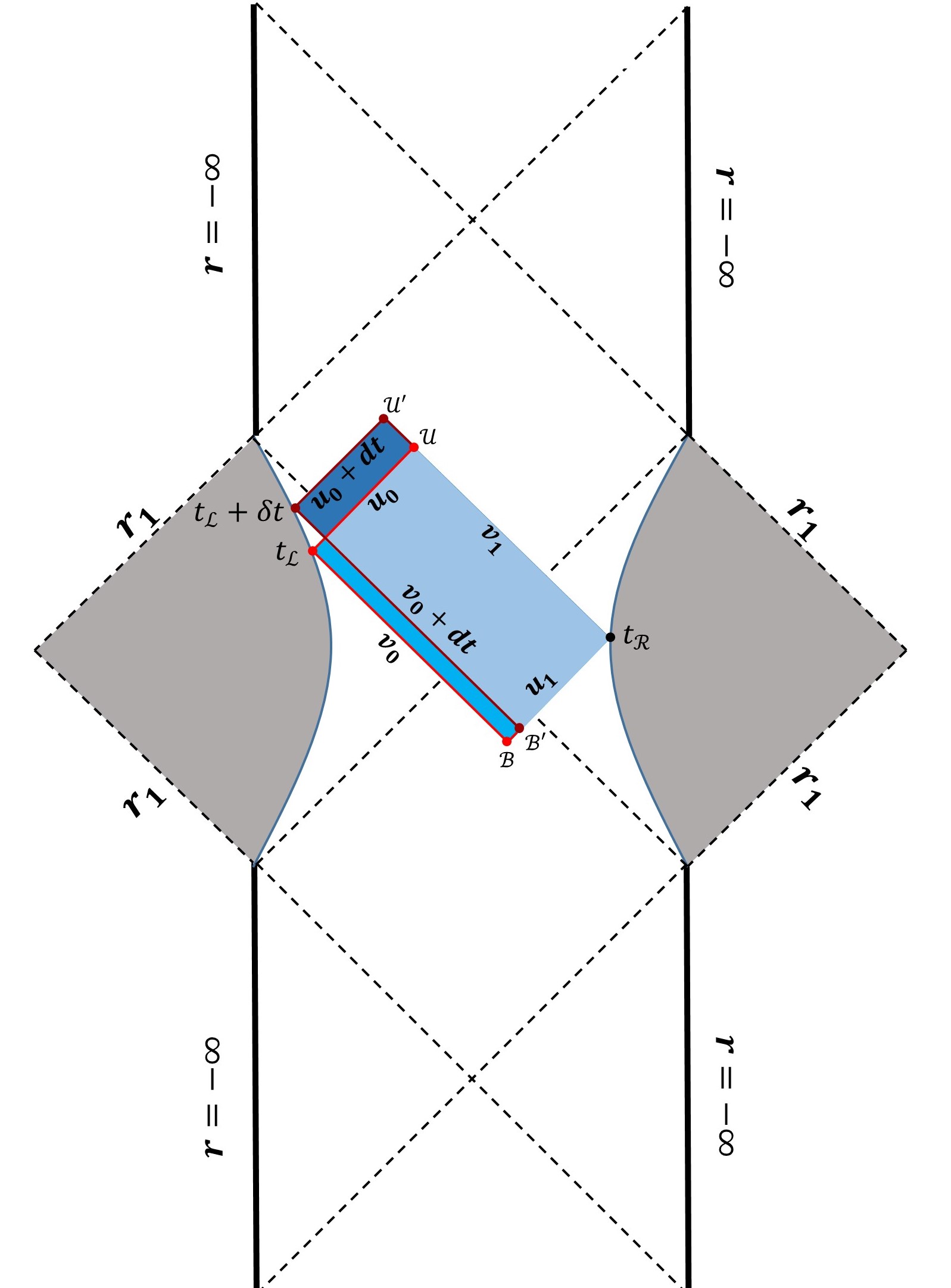}\\
\caption{\textit{Left panel}\label{fig:JTlike}: Penrose diagram for the JT-like black hole (\ref{JT-p}) with outer and inner horizons at $r=r_\pm$ (dashed lines), UV cutoff boundaries at $r=r_b$ (blue lines),  and conformal boundaries at $r=\pm\infty$. \textit{Right panel}: The change of WDW patches from the one anchored on $t_L$ and $t_R$ (pale blue and bright blue regions) to the one anchored on $t_L+\delta t$ and $t_R$ (pale blue and dark blue regions).}
\end{figure}

To resolve the vanishing growth rate of action at late-time for neutral $\mathrm{AdS}_2$ black holes with multiple horizons, a JT-like gravity of form \footnote{A topological term $\frac{1}{2G}\int\sqrt{-g}\mathrm{d}^2xR+\frac{1}{G}\int\sqrt{-h} \mathrm{d}xK$ should be added into Eq.(\ref{eq:CAneutral}) once JT(-like) gravity is regarded as dimensional reduction from a 4D nearly extremal RN black hole(an alternative exact embedding of JT gravity in higher dimension has been studied in \cite{Li:2018omr}). The thermodynamics of black holes with topological term is given in Appendix \ref{top}. It turns out that the topological term makes no contribution to the action growth.}
\begin{align}
\A_{\text{JT-like}}&=\frac{\f_0}{2G}\int_{\M}\mathrm{d}^2x\sqrt{-g}R+\frac{\f_0}{G}\int_{\pd\M}\mathrm{d}x\sqrt{-h}K\nn\\
&+\frac{1}{2G}\int_{\M}\mathrm{d}^2x\sqrt{-g}\Big(\f R+\frac{2\f-3\frac{\f^2}{\f_0}}{L^2}\Big)+\frac{1}{G} \int_{\pd\M}\mathrm{d}x\sqrt{-h}\big(\f K-\frac{\sqrt{\f^2-\frac{\f^3}{\f_0}}}{L}\big)\label{JT-p}
\end{align}
is adopted for illustration, where $\phi/\phi_0\ll1$ for a positive $\f_0$. For a specific solution of form
\begin{align}
\f&=\frac{r}{L},\\
\mathrm{d}s^2&=-f(\f)\mathrm{d}t^2+f(\f)^{-1}\mathrm{d}r^2,\\
f(\f)&=-2GML+\f^2-\frac{\f^3}{\f_0}\\
     &=-2+\f^2-\frac{\f^3}{100}\label{fr-N=2},
\end{align}
with $\f_0=100,~GML=1$, $f(\f)$ exhibits three real roots
\begin{align}
\phi_1&\approx99.9800~,\\
\phi_+&\approx1.4244~,\\
\phi_-&\approx-1.4044~.
\end{align}
Since $\phi_1/\phi_0\approx1$ and here we have assumed $\phi/\phi_0\ll1$ , $\phi_1$ could be neglected. The black hole could be regarded as having double horizons, whose Penrose diagram is shown in Fig.\ref{fig:JTlike} and contains the feature shown in Fig. \ref{fig:CAneutraldouble}. Hence the action growth rate at late-time is also  vanishing. To restore the linear growth of holographic complexity, we first follow the two approaches proposed in \cite{Brown:2018bms,Alishahiha:2018swh}.

\subsubsection{Electromagnetic boundary term}

The JT-like gravity could also be obtained from dimensional reduction of four-dimensional (4D) RN black hole with action of form
\begin{align}
\tI_{\text{RN}}&=\frac{1}{16\pi}\int_{\M}\mathrm{d}^4x\sqrt{-g}\bigg(\frac{1}{\ell^2}R-F_{\m\n}F^{\m\n}\bigg)+\frac{1}{8\pi}\int_{\pd\M}\mathrm{d}^3x\sqrt{-h}\frac{1}{\ell^2}\bigg(K-K_0\bigg)\nn\\&+\frac{1}{4\pi}\int_{\pd\M}\mathrm{d}^3x\sqrt{-h}n_\m F^{\m\n}A_\n,\label{4DRN}
\end{align}
where $\ell^2$ is the 4D Newton constant $G_N$. After adopting an ansatz for a spherically symmetric metric
\begin{align}
\mathrm{d}s^2=\frac{1}{\sqrt{2\Phi}}g_{ij}\mathrm{d}x^i\mathrm{d}x^j+2\ell^2\Phi \mathrm{d}\Omega^2,
\end{align}
the first line in the action (\ref{4DRN}) becomes
\begin{align}
\tI_{2d}=\frac{1}{2}\int \mathrm{d}^2x\sqrt{-g}\bigg(\Phi R+\frac{1}{\ell^2}(2\Phi)^{-\frac{1}{2}}+\frac{\ell^2}{2}(2\Phi)^{\frac{3}{2}}F_{ij}F^{ij}\bigg)+\int \mathrm{d}x\sqrt{-h}\bigg(\Phi K-\frac{1}{\ell}\big(2\Phi\big)^{\frac{1}{4}}\bigg)\label{2DRN}.
\end{align}
After further evaluated at an on-shell electromagnetic field strength, the resulting 2D action reads
\begin{align}
\widetilde{\A}^{2d}_{\oshl}&=\frac{1}{2}\int_{\M} \mathrm{d}^2x\sqrt{-g}\Big(\Phi R+\frac{1}{\ell^2}(2\Phi)^{-\frac{1}{2}}-\frac{Q^2}{\ell^2}(2\Phi)^{-\frac{3}{2}}\Big)+\int_{\pd\M} \mathrm{d}x\sqrt{-h}\bigg(\Phi K-\frac{1}{\ell}\big(2\Phi\big)^{\frac{1}{4}}\bigg),
\end{align}
where the rescaled bulk term $\mathcal{A}\equiv\tilde{\mathcal{A}}/G$ could be rewritten as
\begin{align}
\A_{\bulk}&=\frac{\phi_0}{2G}\int \mathrm{d}^2x\sqrt{-g}R+\frac{1}{2G}\int \mathrm{d}^2x\sqrt{-g}\Big(\phi R+\frac{1}{L^2}V(\phi)\Big)\label{general-Vphi action}\\
&=\frac{\phi_0}{2G}\int \mathrm{d}^2x\sqrt{-g}R+\frac{\f_0}{2G}\int \mathrm{d}^2x\sqrt{-g}\bigg(\frac{\f}{\f_0} R+\frac{1}{L^2}\sum_{n=1}^{N}\frac{(-1)^{n+1}\prod_{k=1}^{n}(2k-1)}{2^{n-2}(n-1)!}\frac{\f^n}{\f_0^n}\bigg)\label{general action}
\end{align}
by expanding $\Phi$ around $\phi_0=Q^2/2$ upto to $N$-th order and abbreviating $L\equiv Q^{\frac{3}{2}}\ell$. For $N=2$, the resulting bulk action
\begin{align}
\A_{\bulk}&=\frac{\f_0}{2G}\int_{\M}\mathrm{d}^2x\sqrt{-g}R+\frac{\f_0}{2G}\int_{\M}\mathrm{d}^2x\sqrt{-g}\Big(\frac{\f}{\f_0} R+\frac{2\frac{\f}{\f_0}-3\frac{\f^2}{\f_0^2}}{L^2}\Big)\label{demost}
\end{align}
coincides with the Eq.(\ref{JT-p}) as expected from dimensional reduction of 4D RN black hole.

The second line in the action (\ref{4DRN}) is a 4D electromagnetic boundary term suggested in \cite{Brown:2018bms}, which induces a 2D boundary electromagnetic boundary term
\begin{align}
 \mathcal{A}^{2d}_\text{em.bdy}&=\frac{\ell^2}{G}\int_{\pd\M} \mathrm{d}x\sqrt{-h}(2\Phi)^{\frac{3}{2}}n_iF^{ij}A_j\nn\\
              &=-\frac{Q^2}{G\ell^2}\int_{\M} \mathrm{d}^2x\sqrt{-g}(2\Phi)^{-\frac{3}{2}}.\label{2DEMBDY}
\end{align}
This should also contribute to the would-be restored JT-like action by $\A_{\text{JT-like}}^{\text{restored}}=\A^{\WDW}_{\text{JT-like}}-\A^{\WDW}_{\text{em.bdy}}$ when evaluated on the WDW patch. Since the first term  $\A^{\WDW}_{\text{JT-like}}$ has vanishing growth rate according to Eq.(\ref{CA0}),
then the growth rate of the restored action at late-time simply reads
\begin{align}
\frac{\mathrm{d}\A_{\text{JT-like}}^{\text{restored}}}{\mathrm{d}t}\bg_{\trf}=-\frac{\mathrm{d}\A^{\WDW}_{\text{em.bdy}}}{\mathrm{d}t}\bg_{\trf}.
\end{align}
After expanded in terms of $\f\sim \f_0$ up to the second order, the contribution from the electromagnetic boundary term reads
\begin{align}
\frac{\mathrm{d}\A^{\WDW}_{\text{em.bdy}}}{\mathrm{d}t}\bg_{\trf}
&=-4\frac{\f_0\f_+}{GL}+2\frac{\f_+^2}{GL}-\frac{3}{2}\frac{\f_+^3}{\f_0GL}+\mathcal{O}(\frac{\f_+^2}{\f_0^2}\f_+^2)
.\end{align}
Therefore, the restored growth rate of $\awdw$ at late-time is
\begin{align}
\frac{\mathrm{d}\A_{\text{JT-like}}^{\text{restored}}}{\mathrm{d}t}\bg_{\trf}=4\frac{\f_0\f_+}{GL}-2\frac{\f_+^2}{GL}+\frac{3}{2}\frac{\f_+^3}{\f_0GL}+\mathcal{O}(\frac{\f_+^2}{\f_0^2}\f_+^2)\label{restored wdw-rate},
\end{align}
which could be expressed in terms of the thermodynamic quantities
\begin{align}
T&=\frac{\f_+}{2\pi L}-\frac{3\fp^2}{4\pi L\f_0}
,~~S=\frac{2\pi(\f_0+\f_+)}{G}
,~~M=\frac{\fp^2}{2GL}-\frac{\fp^3}{2GL\f_0}\label{ST}
\end{align}
as
\begin{align}
\frac{\mathrm{d}\A_{\text{JT-like}}^{\text{restored}}}{\mathrm{d}t}\bg_{\trf}=4ST+G(10M+5S_0T-5ST)=4ST+\mo\big(\frac{\f_+}{\f_0}\f_+^2\big)\label{JTT}
\end{align}
with $S_0\equiv\frac{2\pi\f_0}{G}$. The growth rate (\ref{JTT}) of the JT-like gravity  is identical to the growth rate of the JT gravity given in \cite{Brown:2018bms} on the leading-order. Since the JT-like gravity we demonstrated can be reduced to JT gravity by discarding the sub-leading potential with respect to $\phi\over \phi_0$,i.e., $\frac{\phi^2}{\phi_0^2}$, see \eqref{demost}, the consistency of late-time growth rate indicate that the higher-order correction of dilaton potential in \eqref{JT-p} does not affect the leading order of late-time growth rate but only the sub-leading order.

\subsubsection{UV/IR relation for cutoff surfaces}

\begin{figure}
\centering
\includegraphics[width=0.6\textwidth]{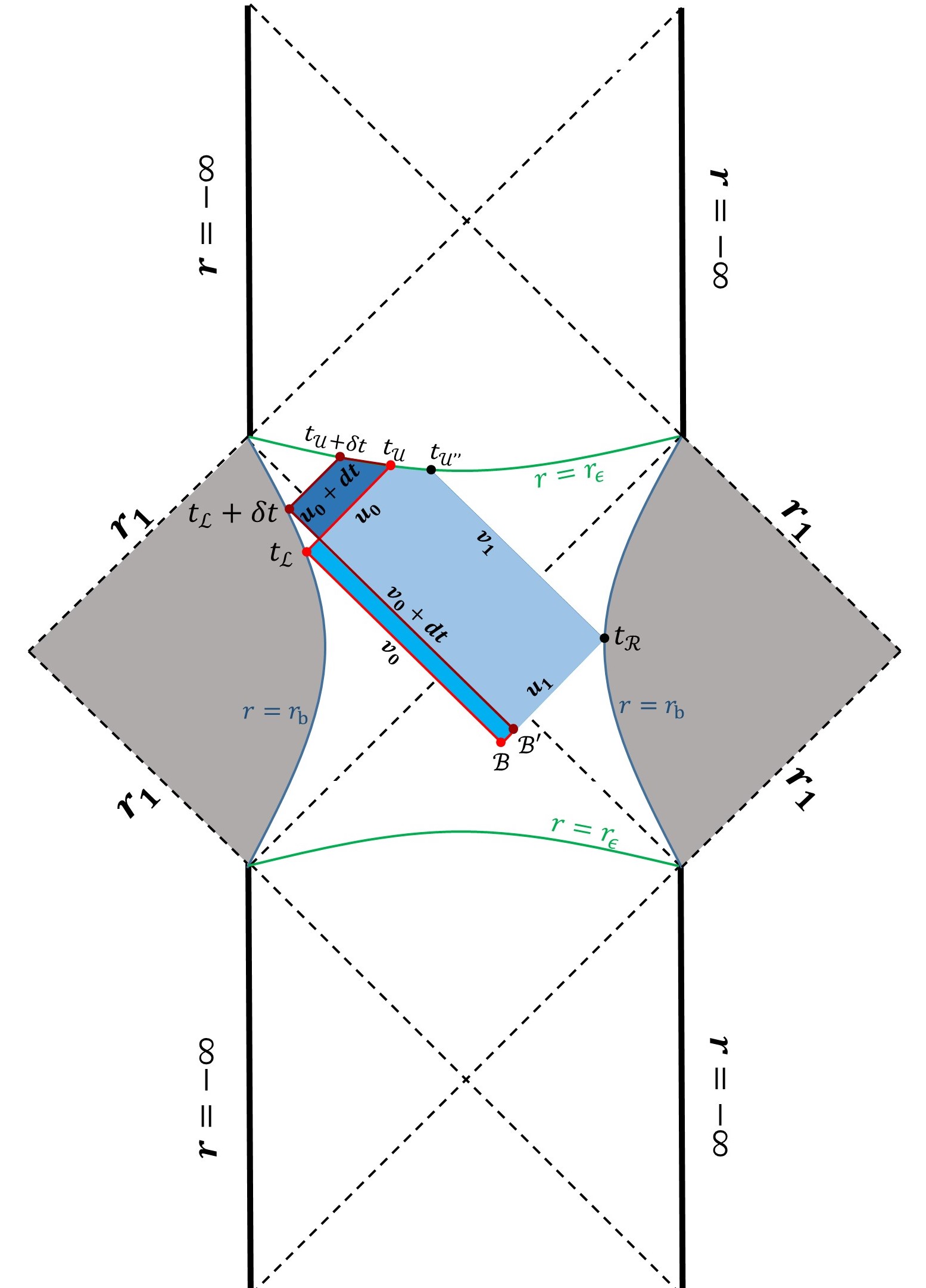}\\
\caption{Penrose diagram for JT-like gravity with both UV cutoff surface $r=r_b$ (blue line) at asymptotic boundary and IR cutoff  surface $r=r_\epsilon$(green line) behind the outer horizon.}\label{fig:UVIR}
\end{figure}

To recover the linear action growth rate at late-time in JT gravity,  \cite{Akhavan:2018wla} proposed an alternative prescription by relating the IR cutoff surface $r=r_\epsilon$ behind the horizon with the UV cutoff surface $r=r_b$ at asymptotic boundary upto the  leading-order
\begin{align}
r_\e=\frac{r_+^3}{r_b^2}\label{cut-off},
\end{align}
and an appropriate counter term \cite{Alishahiha:2018swh} should also be appreciated on the cutoff surface behind the horizon as shown with green line in Fig. \ref{fig:UVIR}. However, there is generally no universal determination for the extra counter term on the $r=r_\e$. Here we introduce an counter term of form $\mathcal{L}^\mathrm{ct}_{\text{JT-like}}[\f_0]=\frac{\sqrt{2}\f_0}{L}$ for the JT-like gravity (\ref{JT-p}), so that the total action
\begin{align}
\A_{\text{JT-like}}^{\text{cut-off}}&=\A_{\text{JT-like}}-\frac{\f_0}{G}\int_{\bdy,r=r_\epsilon}\mathrm{d}x\sqrt{-h}\frac{\sqrt{2}}{L}\label{JT-cc}
\end{align}
exhibits a growth rate
\begin{align}
\frac{\mathrm{d}\A^{\text{cut-off}}_{\text{JT-like}}}{\mathrm{d}t}\bg_{\trf}&=\frac{\sqrt{2}\f_0\f_+}{GL}+\frac{\sqrt{2}\f_+^2}{2GL}+\frac{4-3\sqrt{2}}{4}\frac{\f_+^3}{\f_0GL}+\mo\big(\frac{\f_+^2}{\f_0^2}\f_+^2\big)
\label{fedd}\end{align}
that could be rewritten in terms of thermodynamic quantities as
\begin{align}\label{JT1}
\frac{\mathrm{d}\A_{\text{restored}}}{\mathrm{d}t}\bg_{\trf}=\sqrt{2}ST+\mo\big(\f_+^2\big).
\end{align}
Compared with the growth rate of JT gravity in \cite{Alishahiha:2018swh}
\begin{align}\label{JT2}
\frac{\mathrm{d}\A^{\WDW}_{\JT}}{\mathrm{d}t}\bg_{\trf}=ST+\mathcal{O}(\f_+^2),
\end{align}
the different coefficients in front of $ST$ come from the different counter term  $\mathcal{L}_\mathrm{ct}(\f_0)$ used.

\subsubsection{{Charged dual of neutral black hole}}\label{s1.3}

Apart from the previous two resolutions, we propose here a third solution by  relating the neutral and charged $\mathrm{AdS}_2$ black holes.  Recall that the charged black hole has an action of form
\begin{align}
\A_{\cha}=\frac{1}{2G}\int\mathrm{d}^2x\sqrt{|g|}\Big(\f R+\frac{V(\f)}{L^2}\Big)-\frac{1}{4}\int\mathrm{d}^2x\sqrt{|g|} W(\f)F_{\m\n}F^{\m\n}+\text{GHY term},\label{chargebha}
\end{align}
of which the EOMs and corresponding solution are showed in Eq.(\ref{R-charged}---\ref{a1e1}) and Eq.(\ref{FmnFmn}---\ref{er3}) respectively. As pointed out in \cite{Grumiller:2007ju}, the metric (\ref{re13})  could also be obtained from an uncharged black hole by replacing $V(\f)$ in \eqref{eq:fphi} with an effective potential of the form
\begin{align}
\veff\equiv V(\f)-\frac{GQ^2L^2}{W(\f)},\label{eff}
\end{align}
and the charged and uncharged black hole have the same temperature and Wald entropy. In this sense, there is a dual relation between charged black holes and neutral black holes. Given a charged black hole with dilaton potential $V(\f)$, coupling function $W(\f)$ and electric charge $Q$, the bulk on-shell action is
\begin{align}
\A^{\cha}_{\bulk}&=\frac{1}{2G}\int\mathrm{d}^2x\sqrt{|g|}\Big(\f R+\frac{V(\f)}{L^2}\Big)-\frac{1}{4}\int\mathrm{d}^2x\sqrt{|g|}W(\f)F^2\nn\\
&=\frac{1}{2G}\int\mathrm{d}^2x\sqrt{|g|}\bigg(-\frac{\f V'(\f)}{L^2}+\frac{V(\f)}{L^2}-\frac{GQ^2\f W'(\f)}{W(\f)^2}+\frac{GQ^2}{W(\f)}\bigg),\label{a+}
\end{align}
while the on-shell bulk action of dual neutral black hole with effective potential (\ref{eff}) is
\begin{align}
\A^{\neu}_{\bulk}&=\frac{1}{2G}\int\mathrm{d}^2x\sqrt{|g|}\bigg(\f R+\frac{\veff}{L^2}\bigg)\nn\\
&=\frac{1}{2G}\int\mathrm{d}^2x\sqrt{|g|}\bigg(-\frac{\f V'(\f)}{L^2}+\frac{V(\f)}{L^2}-\frac{GQ^2\f W'(\f)}{W(\f)^2}-\frac{GQ^2}{W(\f)}\bigg)\label{a-},
\end{align}
with flipped sign for $\frac{GQ^2}{W(\f)}$. The difference between (\ref{a-}) and (\ref{a+}) can be rewritten as an additional electric boundary term by using the on-shell electromagnetic field strength (\ref{FmnFmn})
\begin{align}
-\int \frac{Q^2}{W(\f)}d\f=\frac{1}{2}\int dx\sqrt{-h}n_\m W(\f)F^{\m\n}A_{\n}.\label{gnerbd}
\end{align}
For the neutral black hole with the effective potential (\ref{eff}), the action of the corresponding charged black hole is Eq.(\ref{chargebha}) plus Eq.(\ref{gnerbd}). In this sense, the neutral black hole with multiple horizons corresponds to the charged one with varying chemical potential in an ensemble with charged fixed.

To see how the linear growth rate at late-time is restored for neutral $\mathrm{AdS}_2$ black hole with multiple horizons, we start with following action
\begin{align}
\A=\frac{1}{2G}\int\mathrm{d}^2x\sqrt{|g|}\bigg(\Phi R+\frac{(2\Phi)^{-\frac{1}{2}}-G^2\lambda^2(2\Phi)^{-\frac{3}{2}}}{\ell^2}\bigg)+\text{GHY term}\label{an}
,\end{align}
where the JT gravity could be induced from keeping the first order expansion of the dilaton around $\Phi=\f_0=\frac{(G\lambda)^2}{2}$, namely,
\begin{align}
\A\approx\frac{\f_0}{2G}\int\mathrm{d}^2x\sqrt{-g}R+\frac{1}{2G}\int\mathrm{d}^2x\sqrt{-g}\f\Big(R+\frac{2}{L^2}\Big)+\text{High order terms}~,
\end{align}
with $L^2=(G\lambda)^3\ell^2$. The dual charged black hole of (\ref{an}) could be identified by defining
\begin{align}
V(\Phi)=(2\Phi)^{-\frac{1}{2}},~~W(\Phi)=\frac{\ell^2}{G}(2\Phi)^\frac{3}{2},~~Q=\lambda~,
\end{align}
and the corresponding action reads
\begin{align}
\A_{\text{dual}}&= \frac{1}{2G}\int\mathrm{d}^2x\sqrt{|g|} \Big(\Phi R+\frac{1}{\ell^2}(2\Phi)^{-\frac{1}{2}}\Big)
-\frac{1}{4}\int\mathrm{d}^2x\sqrt{|g|}\frac{\ell^2}{G}(2\Phi)^\frac{3}{2}F^{\m\n}F_{\m\n}\nn+\text{GHY term}\\
&+\frac{1}{2}\int dx\sqrt{-h}\frac{\ell^2}{G}(2\Phi)^\frac{3}{2}n_\m F^{\m\n}A_\n \label{RNrduce}.
\end{align}
According to  Eq.(\ref{CAmultiplehorizon}), the late-time action growth rate of the dual charged black hole (\ref{RNrduce}) without the electric boundary term (\ref{gnerbd}) is
\begin{align}
\frac{\mathrm{d}\A_{\text{dual}}}{\mathrm{d}t}\bg_{\trf}&=\m_-\lambda-\m_+\lambda,~~~~\m_\pm=-\frac{\lambda k(\Phi_\pm)}{\a}.\label{adsd}
\end{align}
 After expanding $\Phi_{\pm}$ in (\ref{adsd}) around $\f_0$, $\Phi_\pm=\phi_0\pm\phi_+$, one can obtain the restored growth rate for JT gravity
\begin{align}
\frac{\mathrm{d}\A^{\text{restored}}_{\JT}}{\mathrm{d}t}\bg_{\trf}= \frac{2G\lambda^2}{\a L^2}\f_+~+\mathcal{O}\big(\f_+^2\big)=4ST +\mathcal{O}(T^2).
\end{align}
This is the same as found in \cite{Brown:2018bms}.

\section{CV 1.0}\label{sec:CV1.0}

\begin{figure}
\centering
\includegraphics[width=0.6\textwidth]{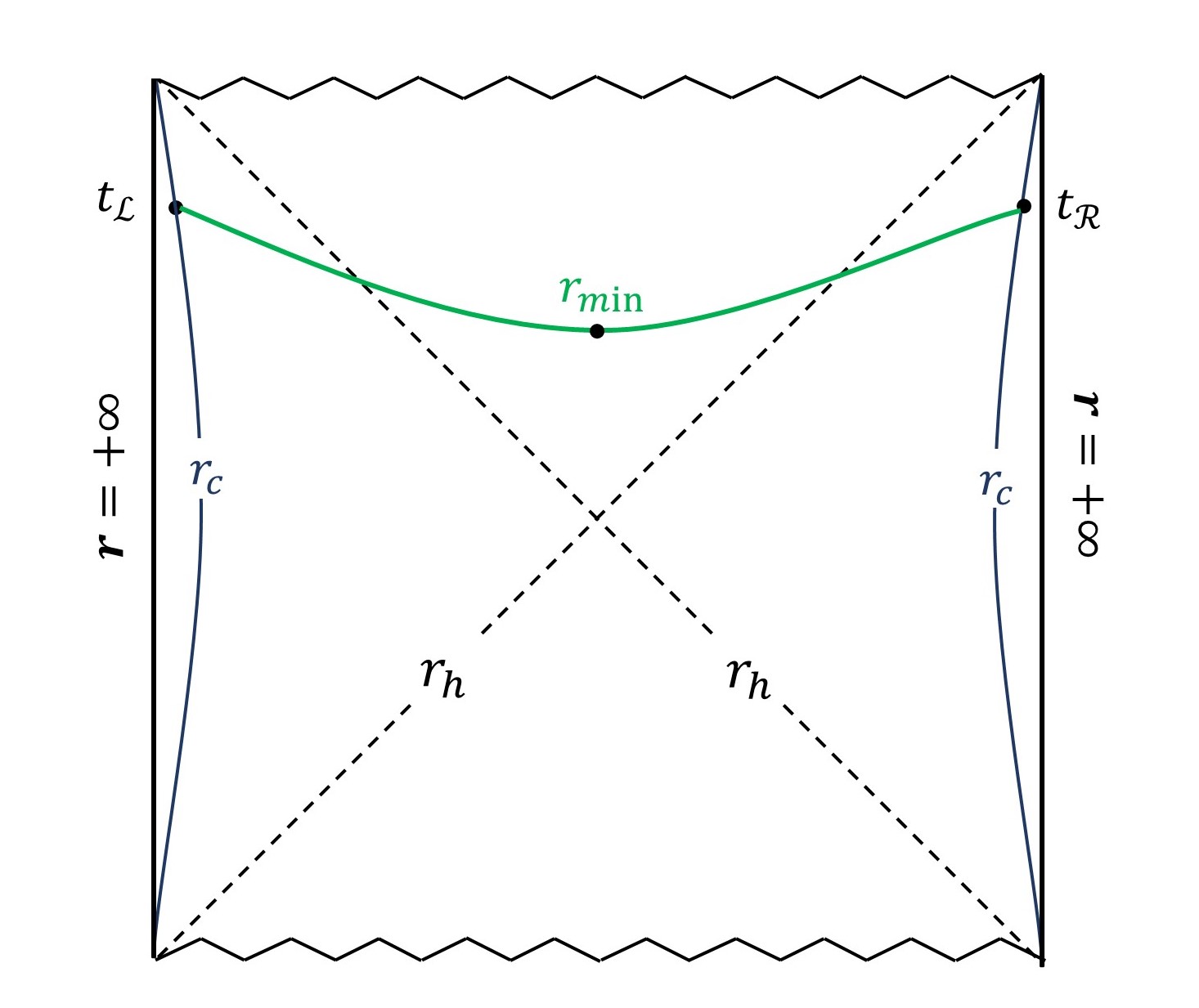}\\
\caption{The extremal surface (green line) anchored at $t_\mathrm{L}$ and $t_\mathrm{R}$ on the boundaries regulated by the UV cutoff surfaces at $r_c$.}\label{fig:CV1.0}
\end{figure}

In this section, we investigate in the context of 2D gravity  the CV 1.0 conjecture \cite{Susskind:2014rva,Stanford:2014jda}, which claims a proportionality between the complexity of the TFD state living on the boundaries and the volume of extremal/maximal time slice anchored at the  boundary times $t_L$ and $t_R$ as shown in Fig. \ref{fig:CV1.0}, namely,
\begin{align}\label{eq:CV1.0}
\mathcal{C}_\mathcal{V}=\max\left[\frac{\mathcal{V}}{G\ell}\right],
\end{align}
where the characteristic scale $\ell$ is set by the $\mathrm{AdS}_2$ radius $L$ for simplicity. Due to the symmetry of TFD state under boundary time shift $t_L\to t_L+\Delta t$, $t_R\to t_R-\Delta t$, the volume should only depend on the total boundary time $t=t_L+t_R$, where a symmetric setup $t_L=t_R$ could be adopted for convenience. Note that the extremal volume behind the horizon exhibits an upper bound \cite{Yang:2019alh}. Following the technique from \cite{Carmi:2017jqz}, we can similarly compute the growth rate of extremal volume for 2D gravity as shown below.

\subsection{Growth rate of extremal volume}

Armed with the in-falling Eddington-Finkelstein coordinate $v=t+r^*(r)$, the metric becomes $\mathrm{d}s^2=-f(r)\mathrm{d}v^2+2\mathrm{d}v\mathrm{d}r$, and hence the volume of the extremal surface parameterized by $r(\lambda)$ and $v(\lambda)$ could be computed by
\begin{align}
\mathcal{V}=\int\mathrm{d}\lambda\mathcal{L}(r, \dot{r}, \dot{v}), \quad \mathcal{L}(r, \dot{r}, \dot{v})\equiv\sqrt{-f(r)\dot{v}^2+2\dot{v}\dot{r}},
\end{align}
where the independence of $\mathcal{L}$ on $v$ gives rise to a conserved quantity $E$ defined by
\begin{align}
E=-\frac{\partial\mathcal{L}}{\partial\dot{v}}=\frac{f\dot{v}-\dot{r}}{\sqrt{-f\dot{v}^2+2\dot{v}\dot{r}}}.
\end{align}
On the other hand, the reparametrization invariance of volume with respect to the choice of $\lambda$ implies that $-f\dot{v}^2+2\dot{v}\dot{r}=1$, therefore, the extremal surface could be determined by
\begin{align}
E&=f(r)\dot{v}-\dot{r},\\
\dot{r}^2&=f(r)+E^2,
\end{align}
and the maximal volume becomes
\begin{align}
\mathcal{V}=2\int_{r_\mathrm{min}}^{r_\mathrm{max}}\frac{\mathrm{d}r}{\dot{r}}=2\int_{r_\mathrm{min}}^{r_\mathrm{max}}\frac{\mathrm{d}r}{\sqrt{f(r)+E^2}},
\end{align}
where $r_\mathrm{min}$ is determined by the condition $\dot{r}=0$, namely, $f(r_\mathrm{min})+E^2=0$. To further evaluate the maximal volume, there is a trick by first noting that
\begin{align}
t_R+r^*(\infty)-r^*(r_\mathrm{min})
&=\int_{v_\mathrm{min}}^{v_\infty}\mathrm{d}v
=\int_{r_\mathrm{min}}^{\infty}\mathrm{d}r\frac{\dot{v}}{\dot{r}}
=\int_{r_\mathrm{min}}^{\infty}\mathrm{d}r\left[\frac{f(r)\dot{v}-\dot{r}}{f(r)\dot{r}}+\frac{1}{f(r)}\right]\\
&=\int_{r_\mathrm{min}}^{\infty}\mathrm{d}r\left[\frac{E}{f(r)\sqrt{f(r)+E^2}}+\frac{1}{f(r)}\right],
\end{align}
then it is easy to see that
\begin{align}
\frac{\mathcal{V}}{2}+E(t_R+r^*_\infty-r^*_\mathrm{min})
&=\int_{r_\mathrm{min}}^{\infty}\mathrm{d}r\left[\frac{1}{\sqrt{f(r)+E^2}}+\frac{E}{f(r)\sqrt{f(r)+E^2}}+\frac{1}{f(r)}\right]\\
&=\int_{r_\mathrm{min}}^{\infty}\mathrm{d}r\left[\frac{\sqrt{f(r)+E^2}}{f(r)}+\frac{E}{f(r)}\right],
\end{align}
which, after taken time derivative with respect to $t_R$, gives rise to
\begin{align}
\frac12\frac{\mathrm{d}\mathcal{V}}{\mathrm{d}t_R}+\frac{\mathrm{d}E}{\mathrm{d}t_R}(t_R+r^*_\infty-r^*_\mathrm{min})+E
&=\int_{r_\mathrm{min}}^{\infty}\mathrm{d}r\frac{\mathrm{d}E}{\mathrm{d}t_R}\left[\frac{E}{f(r)\sqrt{f(r)+E^2}}+\frac{1}{f(r)}\right]\\
&=\frac{\mathrm{d}E}{\mathrm{d}t_R}(t_R+r^*_\infty-r^*_\mathrm{min}),
\end{align}
namely,
\begin{align}
\frac12\frac{\mathrm{d}\mathcal{V}}{\mathrm{d}t_R}=-E.
\end{align}
When written with $t=t_L+t_R=2t_L=2t_R$, one finally arrives at
\begin{align}
\frac{\mathrm{d}\mathcal{V}}{\mathrm{d}t}=-E=\sqrt{-f(r_\mathrm{min})}
\end{align}

\subsection{Growth rate at late-time}

To evaluate the late-time behavior of growth rate of the extremal volume, one has to specify $r_\mathrm{min}$, which is defined by the larger root of $f(r_\mathrm{min})+E^2=0$ with two positive roots meeting at the extremum of $f(r)$ at late-time.

\paragraph{Neutral cases} For neutral black hole, $f(r)=-2GM/\alpha+j(\phi)$, hence $f'(r_\mathrm{min})=0$ leads to $j'(\phi_\mathrm{min})=0$, namely, $V(\phi_\mathrm{min})=0$. For one of the particular choice of potential \eqref{eq:CAneutralV}, $V(\phi)=2\phi+B$, this gives rise to $\phi_\mathrm{min}=-B/2$. After inserting the definition of
\begin{align}
j(\phi)=\int\mathrm{d}\phi\frac{V(\phi)}{\alpha^2L^2}=\frac{\phi^2+B\phi}{\alpha^2L^2},
\end{align}
the growth rate at late-time becomes
\begin{align}
\frac{\mathrm{d}\mathcal{V}}{\mathrm{d}t}\bigg|_{t\to\infty}&=\sqrt{\frac{2GM}{\alpha}-j(\phi_\mathrm{min})}=\sqrt{\frac{2GM}{\alpha}+\frac{B^2}{4\alpha^2L^2}}\nn\\
&=\sqrt{\frac{\phi_+^2+B\phi_+}{\alpha^2L^2}+\frac{B^2}{4\alpha^2L^2}}=\frac{2\phi_++B}{2\alpha L}\equiv2\pi LT,
\end{align}
where in the second line we have used $f(\phi_+)=0$ and $T=\frac{2\phi_++B}{4\pi\alpha L^2}$. This coincides with the result  \cite{Brown:2018bms} from JT gravity when setting  $B=0$.

\paragraph{Charged cases}  For charged black hole, $f'(r_\mathrm{min})=0$ leads to $j'(\phi_\mathrm{min})=GQ^2k'(\phi_\mathrm{min})/\alpha^2$ by noting that
\begin{align}
f(r)=-\frac{2GM}{\alpha}+j(\phi)-\frac{GQ^2}{\alpha^2}k(\phi), \quad j'(\phi)=\frac{V(\phi)}{\alpha^2L^2}, \quad k'(\phi)=\frac{1}{W(\phi)},
\end{align}
namely, $V(\phi_\mathrm{min})=GQ^2L^2/W(\phi_\mathrm{min})$. For our particular choice, $V(\phi)=2\phi$, $W(\phi)=A$, this gives rise to $\phi_\mathrm{min}=GQ^2L^2/(2A)$. Since $j(\phi)=\phi^2/(\alpha^2L^2)$ and $k(\phi)=\phi/A$, the growth rate at late-time becomes
\begin{align}
\frac{\mathrm{d}\mathcal{V}}{\mathrm{d}t}\bigg|_{t\to\infty}
&=\sqrt{\frac{2GM}{\alpha}-j(\phi_\mathrm{min})+\frac{GQ^2}{\alpha^2}k(\phi_\mathrm{min})}=\sqrt{\frac{2GM}{\alpha}+\frac{G^2Q^4L^2}{4\alpha^2A^2}}\nn\\
&=\sqrt{\frac{\phi_+^2}{\alpha^2L^2}-\frac{GQ^2}{\alpha^2}\frac{\phi_+}{A}+\frac{G^2Q^4L^2}{4\alpha^2A^2}}=\frac{\phi_+}{\alpha L}-\frac{GQ^2L}{2\alpha A}\equiv2\pi LT,
\end{align}
where in the second line we have used $f(\phi_+)=0$ and $T=\frac{1}{4\pi\alpha L^2}\left(2\phi_+-\frac{GQ^2L^2}{A}\right)$.

\section{CV 2.0}\label{sec:CV2.0}

We next turn to CV 2.0 conjecture proposed in \cite{Couch:2016exn} that  the holographic complexity of eternal black hole should be proportional to the spacetime volume of the whole WDW patch
\begin{align}
C=\frac{PV_{\WDW}}{\hbar},
\end{align}
of which the late-time limit certainly approaches the product of thermodynamics pressure $P$ and thermodynamic volume $V_\mathrm{th}$
\begin{align}
\lim\limits_{t\rightarrow+\infty}\frac{\mathrm{d}\mathcal{C}}{\mathrm{d}t}=\frac{P\vth}{\hbar}\label{2.0growthrate}
\end{align}
for AdS black hole with a single horizon, or
\begin{align}
\lim\limits_{\trf}\frac{\mathrm{d}\mathcal{C}}{\mathrm{d}t}=\frac{P\big(\vth^+-\vth^-\big)}{\hbar}\label{2.0growthrate2}
\end{align}
for AdS black hole with multiple horizons. Here $\vth^{+}$ and $\vth^-$ is the thermodynamic volume defined by the outer horizon $r_+$ and inner horizon $r_-$, respectively. Eq.(\ref{2.0growthrate}) and Eq.(\ref{2.0growthrate2}) obey the Lloyd bound in many cases as shown in \cite{Couch:2016exn}\cite{An:2018dbz}\cite{Sun:2019yps}. In this section, we would like to investigate CV 2.0  for eternal $\mathrm{AdS}_2$ black holes.

\subsection{Growth rate at late-time}

The WDW patch in the 2D AdS black holes with single horizon and multiple horizons as shown in Fig. \ref{fig:CAneutralsingle} and the Fig. \ref{fig:CAneutraldouble}, respectively.

\paragraph{Single-horizon}

For 2D AdS black hole with a single horizon, the change in spacetime volume of WDW patch could be directly computed by
\begin{align}
\d V_{\WDW}&=\d V_{\text{dark blue}}-\d V_{\text{bright blue}}\nn\\
&=\int_{u_0}^{u_0+\d t}\mathrm{d}u\int^{\rho(\frac{u-(v_0+\d t)}{2})}_{\e}\mathrm{d}r
-\int^{v_0+\d t}_{v_0}\mathrm{d}v\int^{\rho(\frac{u_0-v}{2})}_{\rho(\frac{u_1-v}{2})}\mathrm{d}r\nn\\
&=\d t(\rb-\rcm)+\mathcal{O}(\d t),
\end{align}
of which the late-time limit under  $t\rightarrow+\infty$ , $\rb\rightarrow r_h$ and $\rcm\rightarrow 0$ gives rise to a growth rate as
\begin{align}
\frac{\mathrm{d}V_{\WDW}}{\mathrm{d}t}\bg_{\trf}=r_{h}.
\end{align}

\paragraph{Multiple horizons}

For 2D AdS black hole with multiple horizons, the change in spacetime volume of WDW patch could be directly computed by
\begin{align}
\d V_{\WDW}&=V_{\text{dark blue}}-V_{\text{bright blue}}\nn\\
&=\int_{u_0}^{u_0+\d t}\mathrm{d}u\int^{\rho(\frac{u-v_0-\d t}{2})}_{\r(\frac{u-v_1}{2})}\mathrm{d}r-\int_{v_0}^{v_0+\d t}\mathrm{d}v\int_{\r(\frac{u_1-v}{2})}^{\r(\frac{u_0-v}{2})}\mathrm{d}r\nn\\
&=\d t(\rb-\rc)+\mathcal{O}(\d t),
\end{align}
of which the late-time limit under $t\rightarrow+\infty$ , $\rb\rightarrow r_+$  and $\rc\rightarrow r_-$ gives rise to a growth rate as
\begin{align}
\frac{\mathrm{d} V_{\WDW}}{\mathrm{d} t}\bg_{\trf}=r_+-r_-~.
\end{align}
In summary, one has
\begin{align}\label{cv2.0longtime}
\frac{\mathrm{d}V_{\WDW}}{\mathrm{d}t}\bigg|_{t\to+\infty}=
\begin{cases}
r_h, & \text{single-horizon black holes},\\
r_+-r_-, &\text{multiple-horizon black holes}.
\end{cases}
\end{align}

\subsection{Thermodynamic volume}\label{subsect4}

To see whether Eq.(\ref{cv2.0longtime}) is compatible to Eq.(\ref{2.0growthrate}) and Eq.(\ref{2.0growthrate2}), one first turns to the  thermodynamic volume of black hole chemistry \cite{Kastor:2009wy,Dolan:2010ha,Kubiznak:2014zwa,Cvetic:2010jb,Kubiznak:2016qmn,Johnson:2014yja,Zeng:2016aly}, which treats the cosmological constant $\Lambda$ as the thermodynamic pressure \cite{Kastor:2009wy,Dolan:2010ha,Kubiznak:2014zwa,Cvetic:2010jb} so that the thermodynamic volume $\vth$ plays a role in the extended first law of black hole thermodynamics
\begin{align}
\mathrm{d}M=T\mathrm{d}S+\m\mathrm{d}Q+\vth \mathrm{d}P+...~.
\end{align}
Following the works \cite{Kastor:2009wy,Frassino:2015oca},  we will identify the late-time growth rate of WDW volume with the thermodynamic volume for JT and JT-like (JT+constant potential) black holes.

Before that, there is a subtlety in the definition of entropy in 2D gravity.  As  argued in \cite{Frassino:2015oca} that the Smarr relation for 2D neutral spinless black holes should be modified as
\begin{align}\label{sm2d}
G_2M=2TS_2-2\vth P_2,
\end{align}
where the entropy defined by \cite{Frassino:2015oca}
\begin{align}
S_2=\lim\limits_{D\rightarrow2}S_D=\lim\limits_{D\rightarrow2}\frac{w_{D-2}}{4}\Big(\frac{r_+}{\ell_P}\Big)^{D-2}=\lim\limits_{D\rightarrow2}\Big(\frac{1}{2}+\frac{D-2}{2}\widetilde{S}_{BH}\Big)\label{s2}
\end{align}
seems to have $S_2=\frac{1}{2}$ for the black holes having two horizon points as its ``profile boundary'' with area $A_{\bdy}=1$ for each point. However, as argued in \cite{Grumiller:2007ju}, it is reasonable to redefine
\begin{align}
S_2=\lim\limits_{D\rightarrow2}\Big(\frac{1}{4}+\frac{D-2}{4}\widetilde{S}_{BH}\Big)\label{ts2}
\end{align}
for black hole with only one horizon as its ``profile boundary''.

\subsubsection{JT black hole}

For locally $\AdS_2$ spacetime, like JT gravity,  Eq.(\ref{sm2d}) can be derived from  Komar integral relation \cite{Bazanski:1990qd,Kastor:2008xb}
\begin{align}
\int_{\pd\Sigma}\mathrm{d}S_{ab}\Big(\nabla^a\xi^b-\dl w^{ab}\Big)=0\label{komar},
\end{align}
where $\dl=1/L^2$ is the cosmological constant for 2D spacetime, $\pd\Sigma$ is the  boundary of the spacelike hyper-surface $\Sigma$, $\mathrm{d}S_{ab}=\mathrm{d}S\cdot2[\hat{n}_an_b]$, $n_b$ and $\hna$ is the unit normal to $\Sigma$ and $\pd\S$, respectively, and $w^{ab}$ is the so called killing potential
\begin{align}
\xi^b=\nabla_a w^{ab}.\label{kp}
\end{align}
According to \cite{Frassino:2015oca}\cite{Kastor:2009wy},  Eq.(\ref{komar}) is equal to
\begin{align}
\frac{\hna n_b}{4\pi}\Big(\nabla^a\xi^b-\dl w^{ab}_{\ads}\Big)\bg_{\rrf}=\frac{\hna n_b}{4\pi}\nabla^a\xi^b\bg_{r=r_+}+\frac{\dl}{4\pi}\big(\hna n_bw^{ab}\bg^{+\infty}_{r_+}-\hna n_bw^{ab}_{\ads}\bg_{\rrf}\big),\label{car}
\end{align}
where $w^{ab}_{\ads}$ is the Killing potential for pure $\ads_2$, and  Eq.(\ref{car}) could be regarded as Eq.(\ref{sm2d}) after appreciating  Eq.(\ref{ts2}) and the following definitions
\begin{align}
P_2&=\frac{\dl}{8\pi},\label{P}\\
M&=\frac{1}{4\pi}\hna n_b\Big(\nabla^a\xi^b-\dl w^{ab}_{\ads}\Big)\bg_{\rrf},\label{mass1}\\
\vth&=-\Big(\hna n_bw^{ab}\bg^{+\infty}_{r_+}-\hna n_bw^{ab}_{\ads}\bg_{\rrf}\Big)\label{volume1}.
\end{align}
Therefore, the volume of JT gravity is given by evaluating Eq.(\ref{volume1})
\begin{align}
\vth^{\JT+}=r_+,\label{vjt}
\end{align}
then according to \cite{Couch:2016exn}, the volume defined by $r_-$ is
\begin{align}
\vth^{\JT-}=r_-.
\end{align}
which leads to $\lim\limits_{\trf}\frac{\mathrm{d}\mathcal{C}}{\mathrm{d}t}=\frac{P\big(\vth^+-\vth^-\big)}{\hbar}$. The late-time growth rate in JT gravity satisfies Eq. (\ref{2.0growthrate2}).

\subsubsection{JT-like black holes}

\paragraph{Neutral case} For asymptotically $\AdS_2$  neutral  black holes with dilaton potential \eqref{eq:CAneutralV}, the behavior of $R$ is
\begin{align}
\lim\limits_{\f\rightarrow+\infty}R(\f)=\lim\limits_{\f\rightarrow+\infty}-\frac{V'(\f)}{L^2}=-\frac{2}{L^2}.\label{RR}
\end{align}
According to Eq.(\ref{RR}), the Komar integral relation Eq.(\ref{komar}) should be modified as follows
\begin{align}
\int_{\pd\Sigma}\mathrm{d}S r_an_b\nabla^a\xi^b&=\int_{\S}\mathrm{d}Vn_b\frac{-R}{2}\xi^b\nn\\
&=\frac{\Lambda}{2}\int_{\pd\S}\mathrm{d}S\hat{n}_a n_bV'(\f)w^{ab}-\frac{\Lambda}{2}\int_{\S}\mathrm{d}Vn_b(\nabla_aV'(\f))w^{ab}.\label{komar2}
\end{align}
then Eq.(\ref{komar2}) could be rewritten as
\begin{align}
\frac{\hna n_b}{4\pi}\Big(\nabla^a\xi^b-\dl w^{ab}_{\ads}\Big)\bg_{\rrf}&=\frac{\hna n_b}{4\pi}\nabla^a\xi^b\bg_{r=r_+}+\frac{\dl}{4\pi}\bigg(\frac{1}{2}\hna n_bV'(\f)w^{ab}\bg^{+\infty}_{r_+}\nn\\&-\hna n_bw^{ab}_{\ads}\bg_{\rrf}-\frac{1}{2}\int^{+\infty}_{r_+}\mathrm{d}rn_b\pd_aV'(\f)w^{ab}\bigg)~.\label{car2}
\end{align}
With Eq.(\ref{car2}), the thermal volume (\ref{volume1}) here is
\begin{align}
\vth=-\Big(\frac{1}{2}\hna n_bV'(\f)w^{ab}\bg^{+\infty}_{r_+}-\hna n_bw^{ab}_{\ads}\bg_{\rrf}-\frac{1}{2}\int^{\infty}_{r_+}\mathrm{d}r n_b\pd_{a}V'(\f)w^{ab}\Big)~. \label{volume2}
\end{align}
Setting $\xi^a=\big(\frac{\pd}{\pd t}\big)^a$ and $w^{ab}_{\ads}=r$, Eq.(\ref{volume2}) becomes
\begin{align}
\vth^+=\frac{V(\f_+)-B}{2\a}\label{v3}~.
\end{align}
As a consistent check, the thermal volume (\ref{v3}) of JT-like  could reduce to Eq.(\ref{vjt}) by setting $V(\f)=2\f$. One can see that the late-time growth rate in JT+constant potential gravity satisfies  Eq.~(\ref{2.0growthrate2}).

\paragraph{Charged case} For the charged case, we define the effective dilaton potential (\ref{eff}) introduced in Sect.\ref{s1.3} as an ``effective volume''. The potential (\ref{v3}) then becomes the effective volume
\begin{align}
V^{\eff}_{\text{th}}=\frac{V^{\eff}(\f_+)-B}{2\a},\label{v4}
\end{align}
which contains the contribution of $\m Q$ as seen from
\begin{align}
P_2\cdot V^{\eff}_{\text{th}}&=\frac{\dl}{8\pi}\cdot\frac{V^{\eff}(\f_+)-B}{2\a}\nn\\
&=\frac{\dl}{8\pi}\cdot\Big(\frac{V(\f_+)-B-\frac{GQ^2L^2}{W(\f_+)}}{2\a}\Big)\nn\\
&=P_2\cdot{\vth^+}-\frac{GQ^2}{16\pi\a W(\f_+)}.\label{v5}
\end{align}
Here the second term of  Eq.(\ref{v5}) is proportional to $\m Q$ term with $\mu$ defined by the coefficient of the last term in (\ref{v5}), which is an extended version of \cite{Frassino:2015oca}.

\section{Conclusions}\label{sec:conclusion}

In this paper, the complexity growth has been investigated in terms of various holographic complexity conjectures in generic 2D eternal AdS black holes, which are proposed to be dual to the thermal field double states. For CA conjecture in the context of 2D neutral black holes with double event horizons in JT-like gravity when obtained by dimensional reduction from 4D AdS-RN black hole, it should contain an extra contribution from an electromagnetic boundary term that reproduces the linear late-time growth rate instead of the vanishing result without it. This salvation is similar to the CA case in JT gravity as first found in \cite{Brown:2018bms}. A second proposal involving with a relation for the UV/IR cutoff surfaces is also checked to reproduce the non-vanishing growth rate at late-time. In addition, we propose a third resolution by
explicitly working out the charged dual of a neutral black hole and recurring the vanishing growth, which is consistent with the approach proposed in \cite{Brown:2018bms} for JT gravity.

In the second part of this paper, the late-time growth rates of holographic complexity in terms of CV 1.0 and CV 2.0 are also studied. For CV 1.0, the obtained form of late-time growth rate is universal regardless of neutral or charged black holes with single horizon or multiple horizons. However, it is generally difficult to rewrite it with consistent thermodynamic quantities. For CV2.0 proposed in \cite{Couch:2016exn} that the late-time growth rate of complexity in CV2.0 is equal to the form of $P\vth$ with pressure $P$ associated with the cosmology constant and $\vth$ regarded as the thermal volume, the late-time growth rate for JT-like gravity with constant scalar potential satisfies the form of $\frac{P\big(\vth^+-\vth^-\big)}{\hbar}$ after properly accounting for the thermordynamic first law and Smarr relation.

To close this section, we summarize the main results in the present paper as follows.

\begin{table}[h]
\centering
\caption{Summary of the late-time growth rate of holographic complexity in 2D gravity}\label{table}
\begin{tabular}{|c|c|c|c|c|c|}
\hline\hline
\multicolumn{3}{|c|}{}                      & \multicolumn{3}{c|}{Late-time growth rate:$\lim\limits_{\trf}\frac{\mathrm{d}C}{\mathrm{d}t}$}                                            \\ \hline
\multicolumn{3}{|c|}{Conjecture}            & CA                      & CV1.0                  & CV2.0                 \\ \hline
      &          & \multirow{2}{*}{Neutral} & \multirow{2}{*}{$2M/\pi\hbar$}    & \multirow{8}{*}{$\sqrt{-f(\f_{\min})}/GL$} & \multirow{4}{*}{$\frac{r_h}{8\pi\hbar L^2}^*$} \\
      & Single   &                          &                        &                     &                     \\ \cline{3-4}
      & horizon  & \multirow{2}{*}{U(1)Charged} &
      \multirow{2}{*}{$(2M-\m Q)/\pi\hbar$}  &                     &                     \\
Black &          &                          &                        &                     &                     \\ \cline{2-4} \cline{6-6}
hole  &          & \multirow{2}{*}{Neutral} & \multirow{2}{*}{$\text{Dual to charged case}^{\dagger}$}  &                     & \multirow{4}{*}{$\frac{r_+-r_-}{8\pi\hbar L^2}^*$} \\
      & Multiple  &                          &                        &                     &                     \\ \cline{3-4}
      & horizons & \multirow{2}{*}{U(1)Charged} & \multirow{2}{*}{$(\m_-Q-\m_+Q)/\pi\hbar$} &                     &                     \\
      &          &                          &                        &                     &                     \\ \hline
\end{tabular}
\end{table}
where
\begin{itemize}
  \item [$\dag$]: See Subsect.\ref{s1.3} for more detail,
  \item [*]: See Subsect.\ref{subsect4} for more details. In particular, the late-time growth rate in JT gravity, JT-like gravity with constant scalar potential and their charged versions respectively satisfy the form of $\frac{P\big(\vth^+-\vth^-\big)}{\hbar}$.
\end{itemize}

\section*{Acknowledgements}
The authors are grateful to Xian-Hui Ge, Li Li, Shan-Ming Ruan, Run-Qiu Yang for useful discussions.
SH thanks the Yukawa Institute for Theoretical Physics at Kyoto University.
Discussions during the workshop YITP-T-19-03 ``Quantum Information and String Theory 2019''
were useful to complete this work. RGC  is supported in part by the National Natural Science Foundation of China Grants Nos.  11690022, 11821505, and 11851302, and by the Strategic Priority Re- search Program of CAS Grant NO. XDB23010500 and No.XDB23030100, and by the Key Research Program of Frontier Sciences of CAS.
SH also would like to appreciate the financial support from Jilin University and Max Planck Partner group. SJW is supported by the postdoctoral scholarship of Tufts University from NSF in part.

\appendix
\appendixpage
\addappheadtotoc

\section{Black hole thermodynamics}\label{app:A}

In this appendix, we derive the thermodynamic quantities for generic 2D-dilaton gravity with or without topological term.

\subsection{Without topological terms}

\subsubsection{Neutral black holes}\label{on-shell action}

We start with the Euclidean action of neutral black holes
\begin{align}
\A_\E&=-\frac{1}{2G}\int_{\mathcal{M}}\mathrm{d}^2x\sqrt{g}\bigg(\phi R+\frac{V(\phi)}{L^{2}}\bigg)-\frac{1}{G}\int_{\pd\mathcal{M}}\mathrm{d}x\sqrt{h}\phi K+\frac{1}{G}\int_{\pd \M}\mathrm{d}x\sqrt{h}\lct^{\neu}(\phi),
\end{align}
where $\M$ is a 2D spacetime region outside of the black hole with boundary $\pd\M$.

The EOMs and corresponding solution are showed in Eq.(\ref{eq:CAneutralEOM}) and Eq.(\ref{eq:fphi}-\ref{eqj}) respectively. Consider that
\begin{align}
{\lim_{\f \to +\infty}} j(\f)=+\infty,\nn
\end{align}
the boundary counter term $\lct^{\neu}(\f)$ should be of form
\begin{align}
\lct^{\neu}(\phi)=\frac{1}{L}\sqrt{\int V(\f)\mathrm{d}\phi}=\a\sqrt{j(\phi)},\label{ct-neu}
\end{align}
as we will see below with correct form of free energy corresponding to the on-shell action consisting of following terms. The bulk part of the on-shell action reads
\begin{align}
\A^\oshl_{\bulk}&=-\frac{1}{2G}\int_{\mathcal{M}}\mathrm{d}^2x\sqrt{g}\bigg(\phi R+\frac{V(\phi)}{L^{2}}\bigg)\nn\\
&=-\frac{\b \a j(\f)}{G}\bigg|^{\f_{max}}_{\f_+}+\b\frac{\phi V(\phi)}{2\a GL^2}\bigg|^{\f_{max}}_{\f_{+}}.
\end{align}
The contribution from GHY term reads
\begin{align}
\A^\oshl_{\GHY}&=-\frac{1}{G}\int_{\pd\mathcal{M}}\mathrm{d}x\sqrt{h}\phi K\nn\\
&=-\b\frac{\phi V(\phi)}{2\a GL^2}\bigg|_{\f=\f_{max}}.
\end{align}
where the extrinsic curvature $K$ on $\pd \M$ is
\begin{align}
K&=\frac{V(\phi)}{2\a L^2\sqrt{f(r)}},
\end{align}
with setting the boundary $\pd \mathcal{M}$ to $r=r_\text{max}$.
The contribution from boundary counter term reads

\begin{align}
\A^\oshl_{\ct}&=\frac{1}{G}\int_{\pd\M}\mathrm{d}x\sqrt{h}\mathcal{L}_\ct^\text{neu}(\phi)\nn\\
&=\frac{\b \a}{G}j(\f)\sqrt{1-\frac{2GM}{\a j(\f)}}\Bigg|_{\f_{max}}\nn\\
&=-\b M+\frac{\b \a}{G}j(\f)\bg_{\f_{max}}+\mathcal{O}(j(\f_{max})^{-1})
\end{align}

After summing over above contributions and letting $\f_{max}\rightarrow+\infty$, the total on-shell action is
\begin{align}
\A^\E_{\oshl}&=-\frac{\b \fp V(\fp)}{2\a GL^2}+\frac{\a\b j(\fp)}{2G},\label{Sneutral}
\end{align}
and the corresponding free energy reads
\begin{align}
\mathcal{F}&=\frac{\A^\E_{\oshl}}{\b}=-\frac{\fp V(\fp)}{2\a GL^2}+\frac{\a j(\fp)}{2G}\label{Fneutral}.
\end{align}
In terms of the black hole temperature and entropy \eqref{eq:CAneutralT}, Eq.(\ref{Fneutral}) could be rewritten as
\begin{align}
\mathcal{F}&=-TS+M,
\end{align}
where $M$ is the ADM energy  in terms of the Hamiltonian analysis of the generic 2D theory as referred in \cite{Gegenberg:1995jy}.
A more rigorous argument has been given in Ref.\cite{Grumiller:2007ju} for the counter term taking the form of (\ref{ct-neu}).

\subsubsection{Charged black holes}\label{app.cha}

We next turn to the Euclidean action of charged black holes
\begin{align}
\A_\E&=-\frac{1}{2G}\int_{\mathcal{M}}\mathrm{d}^2x\sqrt{g}\bigg(\phi R+\frac{V(\phi)}{L^{2}}-\frac{G}{2}W(\phi)F^{2}\bigg)-\frac{1}{G}\int_{\pd\mathcal{M}}\mathrm{d}x\sqrt{h}\phi K\nn\\
&+\frac{1}{G}\int_{\pd \M}\mathrm{d}x\sqrt{h}\lct^{\cha}(\phi),\label{cron}
\end{align}
the EOMs and corresponding solution are showed in Eq.(\ref{R-charged}---\ref{a1e1}) and Eq.(\ref{FmnFmn}---\ref{er3}) respectively. Consider that
\begin{align}
\lim\limits_{\f\rightarrow+\infty} \Big(j(\f)-\frac{GQ^2}{\a^2}k(\phi)\Big)&=+\infty,\nn
\end{align}
then the boundary counter term $\lctc(\f)$ should be of form
\begin{align}
\lctc(\f)=\a\sqrt{j(\f)-\frac{GQ^2}{\a^2}k(\f)}+\frac{GQ^2}{\a}\frac{k(\f)}{\sqrt{|f(\f)|}}\label{je}
\end{align}
to get the correct form of free energy corresponding to the on-shell action. The bulk on-shell action reads
\begin{align}
\A^\E_{\bulk}&=-\frac{1}{2G}\int_{\mathcal{M}}\mathrm{d}^2x\sqrt{g}\bigg(R+\frac{V(\phi)}{L^2}-\frac{G}{2}W(\f)F^2\bigg)\nn\\
&=\frac{\b\f V(\f)}{2G\a L^2}\bg^{\fm}_{\fp}-\frac{\b\a j(\f)}{G}\bg_{\fp}^{\fm}-\frac{\b Q^2\f}{2\a W(\f)}\bg_{\fp}^{\fm}.
\end{align}
The contribution from GHY term is
\begin{align}
\A^\E_{\GHY}&=-\frac{1}{G}\int_{\pd \M}\mathrm{d}x\sqrt{h}\phi K \nn\\
&=-\frac{\b\phi V(\phi)}{2G\a L^2}\bigg|_{\f=\f_{max}}+\frac{\b Q^2\f}{2\a W(\f)}\bigg|_{\f=\f_{max}}.
\end{align}
The contribution from counter term is
\begin{align}
\A^\E_{\ct}&=\frac{1}{G}\int_{\pd\M}\mathrm{d}x\sqrt{h}\lctc(\f)\nn\\
&=-\b M+\frac{\b\a j(\f)}{G}\bigg|_{\f_{max}}.
\end{align}
After summing over above contributions and letting $\phi_{max}\rightarrow+\infty$, the total on-shell action is
\begin{align}
\A^\E_{\oshl}&=\A^\E_{\bulk}+\A^\E_{\GHY}+\A^\E_{\ct}\nn\\
&=-\frac{\b}{2G\a L^2}\phi V \bigg|_{\fp}+\frac{\b Q^2 \phi}{2\a W}\bigg|_{\fp}+\frac{\b \a j(\f)}{G}\bigg|_{\fp}-\b M\label{charonshell},
\end{align}
and the corresponding free energy $\mathcal{F}$ reads
\begin{align}
\mathcal{F}&=\frac{\A_{\oshl}}{\b}=-\frac{1}{2G\a L^2}\fp V(\fp)+\frac{ Q^2 \fp}{2\a W(\fp)}+\frac{Q^2}{\a}k(\fp)+M.
\end{align}
Since the temperature and the entropy  of the black hole are defined by
\begin{align}
T&=\frac{f'(r_{+})}{4\pi}=\frac{1}{4\pi\a L^2}\bigg(V(\fp)-\frac{GL^2Q^2}{W(\fp)}\bigg),\\
S&=-\pd_{T}\mathcal{F}=\frac{2\pi\fp}{G},
\end{align}
the $\mathcal{F}$ could be rewritten as
\begin{align}
\mathcal{F}=-TS+M-\m Q,
\end{align}
where
\begin{align}
 \m=-\frac{Qk(\phi_+)}{\a}\label{chemics}
\end{align}
is the chemical potential.

\subsection{With topological terms}\label{top}

When deriving the JT(-like) gravity action from the 4D RN black hole action, there is an extra topological term
\begin{align}
\A_{\topo}&=\frac{\phi_{0}}{2G}\int_{\mathcal{M}}\mathrm{d}^2x\sqrt{-g}R+\frac{\phi_{0}}{G}\int_{\pd \mathcal{M}}\mathrm{d}x\sqrt{-h}K,\\
\phi_{0}&=\frac{q^2}{2}.
\end{align}
where $q$ is the charge of the corresponding RN black hole. For neutral black holes, the on-shell action contribution from the topological term is
\begin{align}
\A^{\neu}_{\topo}&=-\frac{\phi_0}{2G}\int_{\mathcal{M}}\mathrm{d}^2x\sqrt{g}R-\frac{\phi_0}{G}\int_{\pd\mathcal{M}}\mathrm{d}x\sqrt{h}K\nn\\
&=-\frac{\b\phi_{0}V(\phi_{h})}{2G\a L^2}\label{Stop},
\end{align}
while for charged cases, the topological term is
\begin{align}
\A^{\cha}_{\topo}&=-\frac{\phi_0}{2G}\int_{\mathcal{M}}\mathrm{d}^2x\sqrt{g}R-\frac{\phi_0}{G}\int_{\pd\mathcal{M}}\mathrm{d}x\sqrt{h}K\nn\\
&=-\frac{\b\phi_0V(\phi_{h})}{2G\a L^2}+\frac{\b\phi_0 Q^2}{2\a W(\phi_{h})}\label{topchar}.
\end{align}
We will see below the above definitions reproducing the correct form of free energy.

\subsubsection{Neutral black holes}

For neutral black holes with the on-shell Euclidean action of form
\begin{align}
\A_{\text{total}}=&-\frac{\phi_{0}}{2G}\int_{\mathcal{M}}\mathrm{d}^2x\sqrt{g}R-\frac{\phi_{0}}{G}\int_{\pd \mathcal{M}}\mathrm{d}x\sqrt{h}K\nn\\
&-\frac{1}{2G}\int_{\mathcal{M}}\mathrm{d}^2x\sqrt{g}\bigg(\phi R+\frac{V(\phi)}{L^{2}}\bigg)-\frac{1}{G}\int_{\pd\mathcal{M}}\mathrm{d}x\sqrt{h}\big(\phi K-\lctn(\phi)\big),
\end{align}
the on-shell action $\A_{\total}$ from Eq.(\ref{Stop}) and Eq.(\ref{Sneutral}) reads
\begin{align}
 \A^{\oshl}_{\text{total}}=-\frac{\b\phi_{0}V(\phi_{h})}{2GaL^2}+\frac{a\b j(\phi_h)}{2G}-\b\frac{\phi_{h}V(\phi_h)}{2GaL^2},
\end{align}
then the total free energy $\mathcal{F}_{\total}$ is
\begin{align}
 \mathcal{F}_{\total}&=\frac{\A^{\oshl}_{\total}}{\b}=-\frac{\phi_{0}V(\phi_{h})}{2G\a L^2}+\frac{a j(\phi_h)}{2G}-\frac{\phi_{h}V(\phi_h)}{2G\a L^2}=-ST+M.
\end{align}
where the entropy $S$ is
\begin{align}
 S=\frac{2\pi(\phi_h+\phi_0)}{G}\label{entropy with topterm}.
\end{align}

 \subsubsection{Charged black holes}

 For charged black holes with the on-shell Euclidean action of form
 \begin{align}
\A_{\total}=&-\frac{\phi_{0}}{2G}\int_{\mathcal{M}}\mathrm{d}^2x\sqrt{g}R-\frac{\phi_{0}}{G}\int_{\pd \mathcal{M}}\mathrm{d}x\sqrt{h}K-\frac{1}{2G}\int_{\mathcal{M}}\mathrm{d}^2x\sqrt{g}\bigg(\phi R+\frac{V(\phi)}{L^{2}}\bigg)\nn\\&+\frac{1}{4}\int_{\mathcal{M}}\mathrm{d}^2x\sqrt{g}W(\phi)F^2-\frac{1}{G}\int_{\pd\mathcal{M}}\mathrm{d}x\sqrt{h}\big(\phi K-\lctc(\phi)\big),
\end{align}
the on-shell action from Eq.(\ref{charonshell}) and Eq.(\ref{topchar}) reads
\begin{align}
\A^{\oshl}_{\total}&=-\frac{\b}{2G\a L^2}\big(\phi+\phi_0\big) V \bigg|_{\f_{h}}+\frac{\b Q^2 \big(\phi+\phi_0\big)}{2\a W}\bigg|_{\f_{h}}+\frac{\b \a}{G}j(\phi_{h})-\b M,
\end{align}
then the  free energy$ \mathcal{F}_{\total}$ is
\begin{align}
\mathcal{F}_{\total}&=\frac{\A^{\oshl}_{\total}}{\b}=-ST+M-\m Q,
\end{align}
where the entropy $S$ is
\begin{align}
S=\frac{2\pi(\phi_h+\phi_0)}{G}.
\end{align}

\bibliographystyle{JHEP}
\bibliography{ref-may2}

\end{document}